\documentclass[aps,prd,showpacs,amsmath,floats,floatfix,twocolumn,superscriptaddress,nofootinbib]{revtex4}
\usepackage{graphicx,amssymb,amsmath,amsbsy,mathrsfs}
\usepackage{bm}

\usepackage[usenames]{color}
\usepackage{ulem}
\newcommand{\beq}{\begin{equation}}
\newcommand{\eeq}{\end{equation}}
\newcommand{\beqa}{\begin{eqnarray}}
\newcommand{\eeqa}{\end{eqnarray}}

\newcommand{\Mirr}{M_{\text{irr}}}

\newcommand{\Eadm}{E_{\rm ADM}}
\newcommand{\Padm}{P_{\rm ADM}}
\newcommand{\Jadm}{J_{\rm ADM}}
\newcommand{\SMetric}{g}     % spatial metric
\newcommand{\SRicci}{R}      % spatial Ricci tensor
\newcommand{\SRicciS}{R}     % spatial Ricci scalar
\newcommand{\Lapse}{\alpha}       % physical lapse
\newcommand{\Shift}{\beta}   % physical shift
\newcommand{\ExCurv}{K}      % spatial extrinsic curvature
\newcommand{\TrExCurv}{K}    % trace of spatial extrinsic curvature
\newcommand{\A}{A}           % TraceFree spatial extrinsic curvature
\newcommand{\Weight}{\sigma} % weight function in TT decomposition

\newcommand{\dtime}{\partial_t}   % time derivative
\newcommand{\SCD}{{\nabla\!}}     % spatial covariant derivative
\newcommand{\SLong}[1]{({\mathbb L}{#1})} % longitudinal operator

\newcommand{\SFlatMetric}{f}      % flat spatial metric

\newcommand{\SSpatialNormal}{s} 

\newcommand{\CF}{\psi}              % conformal factor

\newcommand{\CMetric}{{\tilde{g}}}    % conformal spatial metric
\newcommand{\CLapse}{\tilde{\alpha}}     % conformal Lapse
\newcommand{\CWeight}{\tilde{\sigma}} % weight function in conformal TTdecomposition
\newcommand{\dtCMetric}{\tilde{u}}  % d/dt of tracefree spatial metric

\newcommand{\CA}{\tilde{A}}         % Conformal TraceFree extrinsic curvature
\newcommand{\CRicciS}{\tilde{R}}    % conformal Ricci scalar

\newcommand{\CCD}{{\tilde\nabla}\!} % conformal covariant derivative
\newcommand{\CCDu}{{\tilde\nabla}}  % conformal covariant derivative with extra
\newcommand{\CLong}[1]{(\tilde{\mathbb L}{#1})} % Conformal longitudinal operator
\newcommand{\CTwoMetric}{\tilde{h}}         % induced metric in S
\newcommand{\CSpatialNormal}{\tilde{s}}

\newcommand{\OmegaOrbitID}{\Omega_0}  % orbital frequency for initial data

\newcommand{\BoundOut}{\mathcal{B}}
\newcommand{\BoundIn}{\mathcal{S}}

\newcommand{\AH}{\mathcal{H}}

\newcommand{\Slice}{\Sigma}

\newcommand{\OnAH}[1]{{{\mathring{ #1}}}}  % Format AH quantities
\newcommand{\SMM}{\chi}

\newcommand{\HorizonShape}{\OnAH{\SRicciS}}
\newcommand{\ShapeMin}{\min(\HorizonShape)}
\newcommand{\ShapeMax}{\max(\HorizonShape)}

\newcommand{\SpinFromShapeMin}{\TypeSC{\SMM}^{\text{min}}}
\newcommand{\SpinFromShapeMax}{\TypeSC{\SMM}^{\text{max}}}

\newcommand{\Spin}{S} %change to J?
\newcommand{\Area}{A} 
\newcommand{\AKV}{\phi}
\newcommand{\TwoGrad}{D}
\newcommand{\TwoLaplacian}{D^2} %MAYBE use D^2 instead but be careful of D^4
\newcommand{\TwoBiharmonic}{D^4} 
\newcommand{\TwoAreaElement}{\medspace dA}
\newcommand{\AKVExpansion}{\Theta}
\newcommand{\AKVShear}{\sigma}
\newcommand{\TwoMetric}{\OnAH{g}}

\newcommand{\ShearNorm}{\Vert \sigma \Vert^2}

\newcommand{\CoordSep}{d}
\newcommand{\ProperSep}{s}

\newcommand{\HorizonMass}{M}
\newcommand{\rexc}{r_{\rm exc}}

\newcommand{\Conformal}[1]{\tilde{ #1 }}
\newcommand{\CSpin}{\Conformal{\Spin}}
\newcommand{\CMass}{\Conformal{\HorizonMass}}

\newcommand{\Ham}{\mathcal{C}}
\newcommand{\Mom}{\mathcal{C}}
\newcommand{\LTwoNorm}[1]{\left\Vert{ #1 }\right\Vert_{L2}}

\newcommand{\circEvLabel}{{\mbox{SKS-0.93-E3}}}
\newcommand{\plungeEvLabel}{{\mbox{SKS-Headon}}}

\newcommand{\TypeAKV}[1]{#1_{\rm AKV}}
\newcommand{\AKVSMM}{\TypeAKV{\SMM}}
\newcommand{\TypeSC}[1]{#1_{\rm SC}}

\newcommand{\Diff}[1]{\Delta #1}
\newcommand{\DiffSMM}{\Diff{\SMM}}
\newcommand{\trelax}{t_{\rm relax}}

\newcommand{\JEE}{\varepsilon_J}

\newcommand{\SoTwoMirrMirr}{\zeta}

\newcommand{\Caltech}{\affiliation{Theoretical Astrophysics 130-33,
    California Institute of Technology, Pasadena, CA 91125}}

\newcommand{\Cornell}{\affiliation{Center for Radiophysics and Space
    Research, Cornell University, Ithaca, NY 14853}}

\begin{document}
\vspace{-2.5cm} 

\title{Binary-black-hole initial data with nearly-extremal spins}

\author{Geoffrey Lovelace} \Cornell\Caltech
\author{Robert Owen} \Cornell\Caltech
\author{Harald P. Pfeiffer} \Caltech
\author{Tony Chu} \Caltech

\date{\today}

\begin{abstract}

There is a significant possibility that astrophysical black holes 
with nearly-extremal spins exist. Numerical simulations of such systems 
require suitable initial data. In this paper,  
we examine three methods of constructing
binary-black-hole initial data, focusing on their ability to generate black
holes with nearly-extremal spins: (i) Bowen-York initial data,
including standard puncture data (based on
conformal flatness and Bowen-York extrinsic curvature), (ii) standard
quasi-equilibrium initial data (based on the 
extended-conformal-thin-sandwich equations, 
conformal flatness, and maximal slicing), and (iii)
quasi-equilibrium data based on the superposition of
Kerr-Schild metrics.  We find that the two conformally-flat methods 
(i) and (ii) perform similarly, with spins up to about 0.99 obtainable at the
{\em initial time}.
However, {\em in an evolution}, we expect the spin to quickly relax
to a significantly smaller value around 0.93 as the initial geometry 
relaxes. For
quasi-equilibrium superposed Kerr-Schild (SKS) data [method (iii)], 
we construct initial data with \emph{initial} spins as large as 0.9997.
We evolve SKS data sets with spins of 0.93 and 0.97 and
find that the spin drops by only a few parts in $10^4$ during the
initial relaxation; therefore,  we expect that SKS
initial data will allow evolutions of binary black holes with relaxed spins
above 0.99. Along the way to these conclusions, we also present 
several secondary results: the
power-law coefficients with which the spin of puncture initial data
approaches its maximal possible value; approximate analytic solutions
for large spin puncture data; embedding diagrams for single spinning
black holes in methods (i) and (ii); non-unique solutions for method (ii).  
All of the initial
data sets that we construct contain sub-extremal black holes, and
when we are able to push
the spin of the excision boundary surface into the super-extremal
regime, the excision surface is always enclosed by  a second, sub-extremal 
apparent horizon.  The quasilocal spin is measured by using approximate
rotational Killing vectors, and the spin is also inferred from the extrema of 
the intrinsic scalar curvature of the apparent horizon. 
Both approaches are found to give
consistent results, with the approximate-Killing-vector spin
showing least variation during the initial relaxation.
\end{abstract}

% 04.20.Ex 	Initial value problem, existence and uniqueness of solutions
% 04.25.D- 	Numerical relativity
% 04.25.dg 	Numerical studies of black holes and black-hole binaries
% 02.70.Hm 	Spectral methods
\maketitle

%#########################
\section{Introduction}
%#########################

There is a significant possibility that black holes with nearly-extremal spins 
exist; by ``nearly-extremal'', we mean that the 
spin $\Spin$ and 
mass $\HorizonMass$ of the hole satisfy 
$0.95 \lesssim \Spin/\HorizonMass^2 \lesssim 1$.
Some models of black-hole 
accretion~\cite{VolonteriEtAl:2005,GammieEtAl:2004,Shapiro:2005}
predict that most black holes will have nearly-extremal spins, 
and observational evidence for black holes with nearly-extremal spins 
includes, 
e.g., estimates of black-hole spins in quasars~\cite{WangEtAl:2006} and 
estimates of the spin of a black hole in a certain binary X-ray 
source~\cite{McClintockEtAl:2006}. There is
considerable uncertainty 
about whether black holes do in fact typically have nearly-extremal spins; 
e.g., some models~\cite{KingPringle:2006,KingEtAl:2008,BertiVolonteri:2008} 
of black-hole accretion do not lead to large spins. 
This uncertainty could be reduced by measuring the holes' spins 
directly using gravitational waves.

This prospect of detecting the gravitational waves emitted by colliding
black holes, possibly  with nearly-extremal spins, 
motivates the goal of simulating these spacetimes numerically.
Indeed,
one focus of
intense research has been spinning black hole binaries, 
including
the discovery of dramatic kicks when two
spinning black holes
merge~\cite{Koppitz2007,Campanelli2007,Gonzalez2007b,%
  Herrmann2007,Choi-Kelly-Boggs-etal:2007,%
  Campanelli2007a,Bruegmann-Gonzalez-Hannam-etal:2007,Baker2007,%
  Schnittman2007} as well as some initial exploration of the orbital
dynamics of spinning
binaries~\cite{Campanelli2007b,Campanelli2006d,Campanelli2006c,Herrmann2007c,MarronettiEtAl:2008,
  Berti2007b}. All  of these simulations start from
puncture initial data as introduced by Brandt and
Br{\"u}gmann~\cite{Brandt1997}.

The simplifying assumptions
employed in puncture initial data make it impossible to construct
black holes with spins arbitrarily close to unity.
The numerical value of the fastest obtainable spin  
depends on which dimensionless ratio is chosen to
characterize ``black hole spin.'' 
Often, dimensionless spin is defined based on 
quasilocal properties of the black hole,
\begin{equation}\label{eq:SMMDef}
\SMM:=\frac{S}{M^2},
\end{equation}
where $S$ is taken to be nonnegative and is a suitable quasilocal spin 
(e.g., obtained using
approximate rotational Killing vectors on the apparent horizon 
 as described, for example, in Appendix~\ref{sec:QuasilocalSpin}) and
$M$ is a suitable quasilocal mass. The latter may be obtained from
Christodoulou's formula relating spin, area and mass of a Kerr black
hole,
\begin{equation}\label{eq:GoodSpinNorm}
\HorizonMass^2:=\Mirr^2+\frac{\Spin^2}{4\Mirr^2},
\end{equation}
where we define the irreducible mass in terms of the area $A$ of the
apparent horizon by
$\Mirr:=\sqrt{A/16\pi}$.  

The quantity $\SMM$ is not preserved during an
  evolution.  Specifically, most black hole initial data are not
  exactly in equilibrium, which leads to transients and emission of an
  artificial pulse of gravitational radiation early in numerical
  simulations.  The geometry in the vicinity of the black holes relaxes
  on a time-scale $\trelax$ (typically a few $M$), and during this
  relaxation, the spin changes by 
\begin{equation}\label{eq:DiffSMM}
\DiffSMM
:=  \SMM\left(t=0\right) - \SMM\left(\trelax\right).
\end{equation} 
When constructing a single spinning black hole with
 standard puncture data~\cite{Brandt1997}, for instance,
$\SMM(t=0)\lesssim 0.98$, which seems encouragingly large. However
Dain~\emph{et al.}~\cite{DainEtAl:2008,DainEtAl:2002} evolved standard puncture
data with initial spin close to this limit, and they find that
the spin rapidly drops to $\SMM(\trelax)\approx 0.93$, i.e. 
$\DiffSMM\approx 0.05$.

For single-black-hole spacetimes, another widely used
  dimensionless spin-measure is the ratio of total
  angular momentum\footnote{We define here
$\Jadm$ by an ADM--like
surface integral at infinity; in axisymmetry this definition coincides with the
standard Komar integral for angular momentum (see
Sec.~\ref{sec:PunctureData} for details.)} $\Jadm$ and Arnowitt--Deser--Misner (ADM) energy
  $\Eadm$,
\begin{equation}
\JEE:=\frac{\Jadm}{\Eadm^2}.
\end{equation}
Dain~\emph{et al.}~noted that $\SMM(\trelax)$ is close to $\JEE$ and
explained this result as follows:
the spacetime is axisymmetric, 
which implies that the angular momentum $\Jadm$ is conserved 
and 
that the black hole's spin equals $\Jadm$. 
Moreover, so long as a negligible fraction of the spacetime's energy 
is carried off by the spurious radiation, the hole's quasi-local 
mass will relax to 
a value of $\Eadm$, giving $\SMM\left(\trelax\right)\approx\JEE$.
Thus conformally-flat Bowen-York data cannot be used 
to simulate black holes with nearly-extremal \emph{equilibrium} spins, even 
though the \emph{initial} spins can be made fairly close to $\SMM=1$.

This paper examines three different approaches of constructing 
black hole
initial data with nearly-extremal spin. First, we revisit puncture
initial data and inversion-symmetric Bowen-York initial data.  We show
that for a single, spinning black hole 
at rest, both approaches are
identical, and we determine spin-limits based purely on initial data
more accurately than before:
\begin{equation}\label{eq:BY-limits}
\JEE\le 0.928200,\qquad\SMM(t=0)\le0.9837.
\end{equation} 
We show that the limiting values of $\JEE$ and $\SMM(t=0)$ 
are approached as power-laws 
of the spin-parameter (curiously, with different powers).
We furthermore give 
 insight into the geometric structure of
  these high-spin Bowen-York initial data sets 
through numerical study and approximate analytical solutions 
and find that a cylindrical
  throat forms which lengthens logarithmically with the spin-parameter.

Second, we investigate the high-spin limit of another popular
approach of constructing initial data, the quasi-equilibrium
formalism~\cite{Gourgoulhon2001,Grandclement2002,Cook2002,Cook2004,Caudill-etal:2006} based on the
conformal thin sandwich equations~\cite{York1999,Pfeiffer2003b}.  
For the standard choices of conformal flatness and maximal slicing, 
we are able to construct initial data with spins somewhat
{\em larger} than the standard Bowen-York limits given in 
Eq.~(\ref{eq:BY-limits}):
\begin{equation}
\JEE\lesssim 0.94,\quad \SMM(t=0)\lesssim 0.99.
\end{equation} 
Once again $\JEE$ is much lower than $\SMM(t=0)$, which suggests that
these data sets lead to equilibrium spins of approximate magnitude
$\SMM\approx 0.94$.  Interestingly, these families of
initial data are found to exhibit non-unique
solutions~\cite{Pfeiffer-York:2005,Baumgarte2007,Walsh2007}, 
and the largest spins are obtained along the upper branch.

The third approach also utilizes the quasi-equilibrium
formalism~\cite{Gourgoulhon2001,Grandclement2002,Cook2002,Cook2004,Caudill-etal:2006}, but this
time we make use of the freedom to chose an arbitrary background data.
Specifically, we choose background data as a superposition 
of two Kerr-Schild metrics.  This approach is based on 
the original proposal of Matzner and
collaborators~\cite{Matzner1999,Marronetti-Matzner:2000} and
was first carried over into the conformal thin sandwich equations in 
Ref.~\cite{Thesis:Lovelace}; 
also, background data consisting of 
a single, non-spinning Kerr-Schild black hole was used to construct 
initial data for a black-hole--neutron-star binary in 
Ref.~\cite{TaniguchiEtAl:2006}.
For single black holes, this data simply reduces to the analytical
Kerr solution.
For binary black holes, we construct 
initial data with spins as
large as
\begin{eqnarray}
\SMM(t=0)  =  0.9997.
\end{eqnarray}
We also present evolutions, demonstrating that our rapidly-spinning
initial data sets remain rapidly-spinning after the numerical
evolution relaxes.  In particular, we evolve an orbiting binary with
$\SMM(t=0)=0.9275$ and a head-on merger with $\SMM(t=0)=0.9701$. 
In both cases, 
$\left|\DiffSMM/\SMM(t=0)\right|$ is significantly smaller than $10^{-3}$.
We conclude that the conformally-curved SKS initial data we present 
in this paper, in contrast with 
conformally-flat Bowen-York data, 
is suitable for simulating binary black holes with nearly-extremal
spins.

Throughout the paper, we use two different techniques to measure the
dimensionless spin of black holes, 
which are described in the appendices. 
The first (Appendix~\ref{sec:QuasilocalSpin})
technique
uses the standard surface-integral based on an 
 approximate rotational Killing vector 
of the apparent horizon. We compute the approximate Killing vector with a
variation of the technique introduced by Cook and
Whiting~\cite{Cook2007}, extended with new normalization conditions of
the approximate Killing 
vector, and we denote the resulting spin ``AKV spin'', $\AKVSMM$.
The second approach (Appendix~\ref{sec:SpinFromShape}) is 
based on the shape of the
horizon in the form of its scalar curvature; specifically,
the spin magnitudes are
inferred from the minimum and maximum of the 
intrinsic Ricci scalar curvature of the horizon. We call the spin inferred 
in this way the ``scalar curvature spin,'' and we label the spin 
magnitudes inferred from the scalar curvature minimum and maximum as
$\SpinFromShapeMin$ and $\SpinFromShapeMax$, respectively. 
Typically, binary-black-hole initial data produces 
holes that are initially 
\emph{not} in equilibrium. Therefore, we use only the AKV spin to 
measure the \emph{initial} black hole spin 
(Secs. \ref{sec:SBHdata}--\ref{sec:BBHdata}.) We use both the AKV 
and the scalar-curvature spin when we measure 
the spin after 
the holes have relaxed to equilibrium (Sec.~\ref{sec:Evolutions}).

We also monitor whether any of the constructed initial data
  sets have super-extremal spins, as this may shed light, for example, on the
  cosmic censorship conjecture. When using the Christodoulou
 formula [Eq.~(\ref{eq:GoodSpinNorm})] to define $M$, the quasilocal
dimensionless spin $\SMM$ 
is {\em by definition}
bounded~\cite{BoothFairhurst:2008}, $\SMM \le 1$. 
This can be seen
most easily by introducing 
the parameter $\SoTwoMirrMirr$, defined as
\begin{equation}\label{eq:qDef}
\SoTwoMirrMirr:=\frac{S}{2\Mirr^2},
\end{equation}
and then rewriting $\SMM$ as
\begin{equation}\label{eq:qFormula}
\SMM=1-\frac{(1-\SoTwoMirrMirr)^2}{1+\SoTwoMirrMirr^2}.
\end{equation} 

The ratio $\SMM$ is therefore not useful to diagnose super-extremal
black holes. A more suitable diagnostic is found in the parameter 
$\SoTwoMirrMirr$.
For Kerr black holes, the first term on the right-hand-side of
Eq.~(\ref{eq:GoodSpinNorm}) is always smaller or equal to the
second, with equality only for extremal spin; i.e.,
$\SoTwoMirrMirr \leq 1$, with equality for 
extremal spin.
This motivates an
alternative definition of 
extremality~\cite{BoothFairhurst:2008}: 
a black hole is said to be
superextremal if the second term in Eq.~(\ref{eq:GoodSpinNorm}) is
larger than the first one, i.e. if $\SoTwoMirrMirr > 1$.  
In this paper, we monitor $\SoTwoMirrMirr$, which we call the 
spin-extremality parameter, along with the dimensionless spin $\SMM$.
We find instances where $\SoTwoMirrMirr$ exceeds
unity.  Before this happens, however, 
a larger, subextremal ($\SoTwoMirrMirr <1$) 
apparent horizon appears, enclosing the smaller, 
superextremal horizon
(Sec.~\ref{sec:SKSBBHID}, Fig.~\ref{fig:IDPlot_BBH_SKS_Maximal}).

This paper is organized as follows.  
Section~\ref{sec:IDF} summarizes the various formalisms that we use 
to construct initial data. Section~\ref{sec:SBHdata} investigates
 single 
black hole initial data, followed by the construction of binary-black-hole 
initial data in Sec.~\ref{sec:BBHdata}.
Section~\ref{sec:Evolutions} 
presents binary black hole
evolutions that show the good properties of superposed Kerr-Schild
data, and the various spin-diagnostics.  We summarize and discuss our results 
in
Sec.~\ref{sec:conclusions}. 
Finally, Appendix~\ref{sec:spin} 
and Appendix~\ref{sec:SpinFromShape}
present our techniques to define black hole spin.

\section{Initial data formalism}
\label{sec:IDF}
Before constructing initial data for rapidly-spinning single 
(Sec.~\ref{sec:SBHdata}) and binary (Sec.~\ref{sec:BBHdata}) black holes, we 
first summarize the initial data formalisms we will use. 
After laying some general groundwork in Sec.~\ref{sec:extr-curv-decomp}, 
we describe Bowen-York initial data (including puncture 
initial data) in Sec.~\ref{sec:PunctureData} 
and 
quasi-equilibrium extended-conformal-thin-sandwich data 
in Sec.~\ref{sec:QEXCTS}.

\subsection{Extrinsic curvature decomposition}
\label{sec:extr-curv-decomp}
Initial data sets for Einstein's equations 
are given on a spatial hypersurface $\Slice$ and 
must satisfy the constraint equations
\begin{align}
\label{eq:Ham}
\SRicciS + \TrExCurv^2 - \ExCurv_{ij}\ExCurv^{ij} & = 0,\\
\label{eq:Mom}
\nabla_j\left(\ExCurv^{ij}-\SMetric^{ij}\TrExCurv\right) & = 0.
\end{align}
Here, $\SMetric_{ij}$ is the induced metric of the slice $\Slice$,
with covariant derivative $\nabla_i$, 
$\SRicciS:=\SMetric^{ij}\SRicci_{ij}$ denotes the trace of the
Ricci-tensor $\SRicci_{ij}$, and 
$\ExCurv_{ij}$ denotes the
extrinsic curvature of the slice $\Sigma$ as embedded into the
space-time manifold ${\cal M}$.

The constraint equations~(\ref{eq:Ham}) and~(\ref{eq:Mom}) 
can be transformed into
elliptic partial differential equations using a conformal
transformation, e.g.~\cite{Pfeiffer2003b}.  One introduces a
conformal metric, $\CMetric_{ij}$ via
\begin{equation}
\label{eq:CMetric}
\SMetric_{ij}=\CF^4\CMetric_{ij},
\end{equation}
with the strictly positive conformal factor $\CF>0$.  
Substituting
Eq.~(\ref{eq:CMetric}) into Eq.~(\ref{eq:Ham}) 
yields an elliptic
equation for $\CF$. One furthermore decomposes the extrinsic curvature
into trace and tracefree part, 
\begin{equation}\label{eq:ExCurvDecomp}
\ExCurv^{ij}=\A^{ij}+\frac{1}{3}\SMetric^{ij}\TrExCurv,
\end{equation}
and splits off a longitudinal part
from the tracefree extrinsic curvature,
\begin{equation}\label{eq:Asplit}
\A^{ij}=\frac{1}{\sigma}{\SLong V}^{ij}+M^{ij}.
\end{equation}
In Eq.~(\ref{eq:Asplit}), $\sigma$ is a strictly positive
weight-function, the longitudinal operator is defined as 
${\SLong
  V}^{ij}=2\nabla^{(i}V^{j)}-\frac{2}{3}\SMetric^{ij}\nabla_kV^k$,
and $M^{ij}$ is symmetric and trace-free\footnote{It is also possible,
  but not necessary, to require that $M^{ij}$ is divergence free.}.
Finally, 
one introduces the conformally scaled quantities
$\sigma=\CF^6\tilde\sigma$, $M^{ij}=\CF^{-10}\tilde M^{ij}$, 
which allows the momentum constraint [Eq.~(\ref{eq:Mom})] 
to be rewritten completely in terms
of conformal quantities: 
\begin{align}
\label{eq:Atilde}
\A^{ij}&=\CF^{-10}\CA^{ij},\\\label{eq:Mtildeij}
\CA^{ij}&=\frac{1}{\tilde\sigma}{\CLong V}^{ij}+\tilde M^{ij}.
\end{align}
The Hamiltonian and momentum constraints then become
\begin{align}\label{eq:Ham2}
\CCD^2\CF-\frac{1}{8}\CRicciS -\frac{1}{12}\TrExCurv^2\CF^5+
\frac{1}{8}\CA_{ij}\CA^{ij}\CF^{-7}&=0,\\
\label{eq:Mom2}
\CCD_j\left(\frac{1}{\tilde\sigma}{\CLong V}^{ij}\right)
-\frac{2}{3}\CF^6\CCD^i\TrExCurv+\CCD_j\tilde M^{ij}&=0.
\end{align}
Given choices for $\tilde M^{ij}$, $\TrExCurv$, $\CMetric_{ij}$ and
$\tilde\sigma$, {\em and also} boundary conditions, 
one can solve
Eqs.~(\ref{eq:Ham2}) and (\ref{eq:Mom2}) for $\CF$ and $V^i$, and then
assemble the (constraint-satisfying) initial data $\SMetric_{ij}$ and
$\ExCurv^{ij}$.  

Many important approaches to construct binary black hole initial data
can be cast in this form.  The various approaches differ in the
choices for the freely specifiable parts and the boundary conditions.
Some choices of free data aim for simplicity, 
such as Bowen-York initial
data.
Other approaches aim to preserve
freedom, resulting in more complicated sets of equations but also
more flexibility to control properties of the resulting initial data.
The quasi-equilibrium extended-conformal-thin-sandwich approach falls into
this second category, and we will exploit precisely its inherent freedom
in choosing the free data to construct black holes with nearly-extremal
spins.

%#####################################
\subsection{Bowen-York initial data}
\label{sec:PunctureData}
%####################################

In this section, we describe two approaches of constructing 
initial data based on 
the well-known Bowen-York extrinsic curvature. 
These two approaches, puncture data and 
inversion-symmetric data, differ 
in how they treat the coordinate singularity at $r=0$;
both can be obtained from
the general
procedure outlined in Sec.~\ref{sec:extr-curv-decomp} 
by setting
$\CWeight\equiv 1$, $\TrExCurv\equiv 0$, $\tilde M^{ij}\equiv 0$ and
by using a conformally flat metric 
\begin{eqnarray}
\CMetric_{ij}=\SFlatMetric_{ij}.
\end{eqnarray}
The momentum constraint [Eq.~(\ref{eq:Mom2})]
then reduces to
$\CCD_j\CLong V^{ij}=0$, which is solved by choosing the analytical
Bowen-York solutions~\cite{bowen79,Bowen-York:1980}. 

The Bowen-York
solutions can be written down most conveniently in Cartesian
coordinates, $\SFlatMetric_{ij}=\delta_{ij}$:
\begin{align}
\label{eq:V-BY-P}
V_P^i &= -\frac{1}{4r}\left[7 P^i + n^i P^kn_k\right],\\
\label{eq:V-BY-S}
V_S^i &= -\frac{1}{r^2}\epsilon^i{}_{lm}S^ln^m,
\end{align}
where $r=(x^i x^j \delta_{ij})^{1/2}$ is the coordinate distance to the
origin and $n^i=x^i/r$ is the coordinate unit vector pointing from
the origin to the point under consideration. The spatially-constant
vectors $P^i$ and $S^i$ parametrize the solutions\footnote{In
  Cartesian coordinates, upper and lower indices are equivalent, so
 index positioning in 
  Eqs.~(\ref{eq:V-BY-P})--(\ref{eq:A-BY-S})
  is unimportant. To find $\CA^{ij}_{P/S}$ in another coordinate
  system, first compute the Cartesian components
  Eqs.~(\ref{eq:V-BY-P})--(\ref{eq:A-BY-S}), and then 
  apply the desired coordinate transformation.}
\begin{align}
\label{eq:A-BY-P}
\CA^{ij}_P &= \frac{3}{2r^2}\left[2 P^{(i} n^{j)} 
   - \left(\delta^{ij}-n^in^j\right)P_kn^k\right],\\
\label{eq:A-BY-S}
\CA^{ij}_S &= \frac{6}{r^3} n_{(i}\epsilon_{j)kl} S^k n^l.
\end{align}

The conformal factor $\CF$ is then
determined by the Hamiltonian 
constraint [Eq.~(\ref{eq:Ham2})], which simplifies to
\begin{equation}
\label{eq:Ham3}
\CCD^2 \CF + \frac{1}{8}\CF^{-7} \CA^{ij}\CA_{ij}=0.
\end{equation}
We would like to recover an asymptotically flat
space; this implies the boundary condition 
$\CF\to 1$ as $r\to\infty$.  

This boundary condition makes it possible to evaluate the linear
ADM-momentum and ADM-like angular momentum
of Bowen-York initial data {\em without}
solving Eq.~(\ref{eq:Ham3}).  These quantities are defined by 
surface integrals at infinity,
\begin{equation}\label{eq:ADM}
J_{(\xi)}  =  \frac{1}{8 \pi}\oint_\infty 
\left(\ExCurv_{ij} - \SMetric_{ij} K\right) \xi^i s^j\TwoAreaElement,\\
\end{equation}
where $s^i$ is the outward-pointing unit-normal to the integration
sphere\footnote{At infinity, the normal to the sphere $s^i$ is
  identical to the coordinate radial unit vector $n^i$.}.  
By letting
$\CF\to 1$ in Eq.~(\ref{eq:Atilde}), one can replace $K_{ij}$ by
$\tilde A_{ij}$ and then evaluate the resulting integrals.  
The choice of
vector $\xi^i$ determines which quantity is computed: For
instance, $\xi=\hat e_x$ corresponds to the x-component of the linear
ADM-momentum, $\xi=\partial_\phi=-x\hat e_y +y\hat e_x$ yields the
z-component of the ADM-like angular
momentum\footnote{\label{footnote:JADM}
As is common 
in the numerical relativity community, we introduce the phrase ``ADM angular 
momentum'' to refer to an angular momentum defined at spatial infinity in the 
manner of the other conserved ADM quantities of asymptotically flat 
spacetimes~\cite{ADM}, despite the fact that (at least to our knowledge), no 
such quantity is widely agreed to rigorously exist in general, due to the 
supertranslation ambiguity that exists in four spacetime dimensions.  For 
recent research on this issue see~\cite{AshtekarEngleSloan2008} and 
references therein.  In the present paper, this subtlety can be ignored, 
because we only compute this quantity in truly axisymmetric spacetimes, 
with $\vec \xi$ the global axisymmetry generator, 
so that $\Jadm$ coincides with the standard Komar integral for angular 
momentum.}.  For Eqs.~(\ref{eq:A-BY-P}) and (\ref{eq:A-BY-S}), 
the results are $\Padm^i=P^i$ and $\Jadm^i=S^i$, respectively.

The ADM energy is given by the expression
\begin{equation}\label{eq:Eadm}
\Eadm  =  \frac{1}{16\pi}\oint_\infty 
\SCD_j\left({\cal G}_i{}^j-\delta_i{}^j{\cal G}\right)s^i\TwoAreaElement,
\end{equation}
where ${\cal G}_{ij}:=\SMetric_{ij}-f_{ij}$, ${\cal G}:={\cal
  G}_{ij}\SMetric^{ij}$. 
For conformal flatness,
Eq.~(\ref{eq:Eadm}) reduces to
\begin{equation}
\Eadm=-\frac{1}{2\pi}\oint_\infty \partial_r\CF \TwoAreaElement.
\end{equation}
The derivative of the conformal factor is known only after
Eq.~(\ref{eq:Ham3}) is solved; 
therefore, in contrast with the linear and angular momenta,
$\Eadm$ can be computed only
after solving the Hamiltonian constraint.

We now turn our attention to inner boundary conditions.  $\CA_P^{ij}$
and $\CA_S^{ij}$ are singular at $r=0$.  This singularity is
interpreted as a second asymptotically flat universe; when solving
Eq.~(\ref{eq:Ham3}), this can be incorporated in two ways:
\begin{itemize}
\item {\bf Inversion Symmetry}: The demand that the solution be
  symmetric under inversion at a sphere with radius $R_{\rm inv}$
  centered on the origin~\cite{Bowen-York:1980} results in a boundary
  condition for $\CF$ at $r=R_{\rm inv}$, namely $\partial\CF/\partial
  r=-\CF/(2R_{\rm inv})$.  The Hamiltonian constraint
  Eq.~(\ref{eq:Ham3}) is solved only in the exterior of the sphere,
  $r\ge R_{\rm inv}$, and the solution in the interior can be
  recovered from inversion symmetry~\cite{Bowen-York:1980}, e.g.
\begin{equation}\label{eq:CF-InversionFormula}
\CF\left(x^i\right) = \frac{R_{\rm inv}}{r}\CF\left(\frac{R_{\rm
    inv}^2}{r^2}x^i\right).
\end{equation}

\item {\bf Puncture data}: One demands~\cite{Brandt1997} the
  appropriate singular behavior of $\CF$ for $r\to 0$ to ensure that
  the second asymptotically flat end is indeed flat.  That is, $\CF$
  must behave as 
\begin{equation}\label{eq:u}
\CF(x^i)=\frac{m_p}{2r}+1+u(x^i)
\end{equation}
for some positive parameter $m_p$ (the ``puncture mass'') and 
function $u(x^i)$ that
is finite and continuous in $\mathbb{R}^3$
and approaches $0$ as $r\to\infty$.  Equation~(\ref{eq:Ham3}) then
implies an equation for $u$
that is finite everywhere and can be solved 
without any inner boundaries:
\begin{equation}\label{eq:Puncture-u-Eqn}
\CCD^2u = -\frac{1}{8}
\frac{\CA_{ij}\CA^{ij}\, r^7}{\left(r+\frac{m_p}{2}+ur\right)^7}.
\end{equation}
The majority of binary black hole simulations use puncture data,
  see, e.g., Refs. [9-23].
\end{itemize}

Both approaches allow specification of multiple black-holes at
different locations, each with different spin and momentum parameters
$S^i$ and $P^i$.  For puncture data this is almost trivial; 
this accounts
for the popularity of puncture data as initial data for black hole
simulations.  In contrast, for inversion-symmetric data, one needs to employ
a rather cumbersome imaging procedure\footnote{Even for a single black
  hole with $P^k\neq 0$, Eq.~(\ref{eq:A-BY-P}) has to be augmented by
  additional terms of ${\cal O}(r^{-4})$ to preserve inversion
  symmetry~\cite{Bowen-York:1980}.} (see e.g. \cite{Cook2000} for details).

For a single spinning black hole at the origin, the extrinsic
curvature $\CA^{ij}_S$ given by Eq.~(\ref{eq:A-BY-S}) is identical for
inversion-symmetric and puncture data.  
For inversion-symmetric
data, the conformal factor has the usual falloff at large radii,
\begin{equation}
\CF(x^i)=1+\frac{\Eadm}{2r}+{\cal O}(r^{-2}),\quad\mbox{as $r\to\infty$}.
\end{equation}
Using Eq.~(\ref{eq:CF-InversionFormula}) we find the behavior of $\CF$ as $r\to 0$:
\begin{align}
\CF(x^i)
%&=\frac{R_{\rm inv}}{r}\left(1+\frac{\Eadm}{2\, R_{\rm inv}^2/r}+{\cal O}(r^2)\right)\nonumber \\
&=\frac{R_{\rm inv}}{r}
+\frac{\Eadm}{2R_{\rm inv}}+{\cal O}(r),\quad\mbox{as $r\to 0$}.
\end{align} 
Comparison with Eq.~(\ref{eq:u}) shows that this is precisely the
desired behavior for puncture data, if one identifies $R_{\rm
  inv}=m_p/2$ and $E/(2R_{\rm inv})=1+u(0)$.  Because puncture data has
a unique solution, it follows that for single spinning black holes,
puncture data and inversion-symmetric data are {\em identical},
provided $m_p=2R_{\rm inv}$.

For inversion-symmetric initial data 
for a single, spinning black hole, it is
well-known~\cite{yorkpiran} that the apparent horizon coincides with
the inversion sphere, $r_{\rm AH}=R_{\rm inv}$.  Therefore, we
conclude that for puncture data 
for a single, spinning black hole, the apparent horizon
is an exact coordinate sphere with radius $r_{\rm AH}=m_p/2$, despite
$\CA^{ij}_S$ and $u(x^i)$ not being spherically symmetric.

%###############################################
\subsection{Quasi-equilibrium 
extended-conformal-thin-sandwich initial data}
%###############################################
\label{sec:QEXCTS}
Another popular approach of constructing binary-black-hole 
initial data is the quasi-equilibrium extended-conformal-thin-sandwich 
(QE-XCTS) formalism~\cite{Cook2002,Cook2004,Caudill-etal:2006,Gourgoulhon2001,Grandclement2002}. Instead of emphasizing the extrinsic curvature,
the conformal thin sandwich formalism~\cite{York1999} emphasizes
the spatial metric $\SMetric_{ij}$
and its {\em time-derivative}.
Nevertheless, it is equivalent~\cite{Pfeiffer2003b} to the extrinsic
curvature decomposition outlined in
Sec.~\ref{sec:extr-curv-decomp}.
The vector $V^i$ is identified with the shift $\Shift^i$, 
\begin{equation}
V^i\equiv \Shift^i,
\end{equation}
and the weight-functions $\Weight$ and $\CWeight$ are identified (up
to a factor $2$) with the lapse and the conformal lapse, respectively,
\begin{equation}
\Weight\equiv2\Lapse,\qquad \CWeight\equiv2\CLapse.
\end{equation}
The tensor $\tilde M_{ij}$ is related to the time-derivative of the
spatial metric, $\dtCMetric_{ij}:=\dtime\CMetric_{ij}$ by 
\begin{equation}
\tilde M_{ij}\equiv\frac{1}{2\CLapse}\dtCMetric_{ij}.
\end{equation} 
Because $M_{ij}$ is trace free 
[Eqs.~(\ref{eq:ExCurvDecomp}) and (\ref{eq:Atilde})--(\ref{eq:Mtildeij})], 
we require $\dtCMetric_{ij}$ to be trace free.

The conformal thin sandwich equations allow control of certain
time-derivatives in the subsequent evolution 
of the constructed initial
data.  If the lapse $\Lapse$ and shift $\Shift^i$ 
from the initial data
are used in the evolution, for instance, 
then the trace-free part of
$\partial_t\SMetric_{ij}$ will be proportional to $\dtCMetric_{ij}$.
Therefore (see Refs.~\cite{Gourgoulhon2001,Cook2002})
\begin{subequations}
\begin{equation}\label{eq:freeTimeDerivsVanisha}
\dtCMetric_{ij}\equiv0
\end{equation} is a preferred choice for 
initial data sets that begin 
nearly in equilibrium, such as
binary black holes quasi-circular orbits.

The evolution equation for $\TrExCurv$ can be used to derive
an elliptic equation for the conformal lapse $\CLapse$ 
(or, equivalently, for $\Lapse\CF$). Upon
specification of 
\begin{equation}\label{eq:freeTimeDerivsVanishb}
\partial_t\TrExCurv\equiv 0, 
\end{equation}
\end{subequations}
this fifth elliptic
equation is to be solved for $\CLapse$ simultaneously with
Eqs.~(\ref{eq:Ham2}) and~(\ref{eq:Mom2}), cf.~\cite{Gourgoulhon2001,Cook2002}.

Our numerical code uses the conformal factor $\CF$, the shift
$\Shift^i$, and the product of lapse and conformal factor
$\Lapse\CF=\CLapse\CF^7$ as independent variables, in order to
simplify the equation for $\partial_t\TrExCurv$.  Thus, the actual
equations being solved take the form 
\begin{subequations}\label{eq:XCTS}
\begin{eqnarray}\label{junkeq:XCTSa}
%\fl 
0 & = & \CCDu^2\CF-\frac{1}{8}\CRicciS\CF-\frac{1}{12}\TrExCurv^2\CF^5 
+\frac{1}{8}\CF^{-7}\CA^{ij}\CA_{ij},\\
%\label{junkeq:XCTSb}
%\fl
0 & = & \CCD_j\Big(\frac{\psi^7}{2(\alpha\psi)}\CLong{\Shift}^{ij}\Big)
-\frac{2}{3}\CF^6\CCDu^i\TrExCurv\nonumber\\
&-&\CCD_j\Big(\frac{\psi^7}{2(\alpha\psi)}\dtCMetric^{ij}\Big),\\
%\fl
0 & = & \CCDu^2(\Lapse\CF)-(\Lapse\CF)\bigg[\frac{\CRicciS}{8}\!+\!\frac{5}
{12}\TrExCurv^4\CF^4\!
+\!\frac{7}{8}\CF^{-8}\CA^{ij}\CA_{ij}\bigg]
%\label{junkeq:Lapse2}
\nonumber\\
&+&\CF^5(\dtime\TrExCurv-\Shift^k\partial_k\TrExCurv),
\label{junkeq:XCTSc}
\end{eqnarray}
with
\begin{equation}
\CA_{ij}=\frac{\CF^7}{2\Lapse\CF}\left(\CLong{\Shift}_{ij}-\dtCMetric_{ij}\right).
\end{equation}
\end{subequations}

These equations can be solved only after
\begin{enumerate}
\item specifying the remaining free data: i.e., the conformal
  metric $\CMetric_{ij}$ and the trace of the extrinsic curvature
  $\TrExCurv$ (we chose already $\dtCMetric_{ij}\equiv0$ and
  $\dtime\TrExCurv\equiv0)$,
\item choosing an inner boundary
$\BoundIn$ which excises the black holes' singularities, and also an outer 
boundary $\BoundOut$, and 
\item choosing boundary conditions for $\CF$, $\Lapse\CF$, and $\Shift^i$ on 
$\BoundOut$ and $\BoundIn$.
\end{enumerate}

The initial data is required to be asymptotically flat, and the outer
boundary $\BoundOut$ is placed at infinity\footnote{In practice,
$\BoundOut$ is
 a sphere with radius
$\gtrsim 10^{9}$ times the coordinate radius of the black-hole horizons.}.
If $\CMetric_{ij}$ is asymptotically flat, the outer boundary conditions are 
then
\begin{subequations}
\begin{eqnarray}\label{eq:OutBCStart}
\CF & = & 1\mbox{ on }\BoundOut,\\
\Lapse \CF & = & 1\mbox{ on }\BoundOut,\\
\Shift^i & = & (\mathbf{\OmegaOrbitID} \times \mathbf{r})^i 
+ \dot{a}_0 r^i\mbox{ on }\BoundOut.\label{eq:OutBCEnd}
\end{eqnarray}
\end{subequations} Here $r^i$ is the coordinate position vector.
The shift boundary condition consists
of a rotation (parametrized by the orbital 
angular velocity $\mathbf{\OmegaOrbitID}$) 
and an expansion (parametrized by $\dot a_0$);
the initial radial velocity is necessary for
reducing orbital
eccentricity in binary-black-hole initial
data~\cite{Pfeiffer-Brown-etal:2007}.

The inner boundary condition on the conformal factor $\CF$ ensures
that the excision surfaces $\BoundIn$ are apparent
horizons~\cite{Cook2002}:
\begin{eqnarray}\label{eq:AH-BC}
\CSpatialNormal^k\partial_k \CF & = & 
-\frac{\CF^{-3}}{8\CLapse}\CSpatialNormal^i\CSpatialNormal^j
\left[\CLong{\Shift}_{ij}-\dtCMetric_{ij}\right]\nonumber\\
&&-\frac{\CF}{4}\,\CTwoMetric^{ij}\CCD_i\CSpatialNormal_j
+\frac{1}{6}\TrExCurv\CF^3 \mbox{ on }\mathcal{S}.
\end{eqnarray} Here $\CSpatialNormal^i:=\CF^2 \SSpatialNormal^i$, 
$\SSpatialNormal^i$ is unit vector normal to $\BoundIn$, and 
$\CTwoMetric_{ij}:=\CMetric_{ij}-\CSpatialNormal_i \CSpatialNormal_j$ is 
the induced conformal 2-metric on $\BoundIn$.

The inner boundary condition on the shift is
\begin{eqnarray}\label{eq:InnerShiftBC}
\Shift^i = \Lapse s^i - \Omega_r \xi^i\mbox{ on }\BoundIn,
\end{eqnarray} 
where $\xi^is_i=0$. The first term on the right-hand-side ensures
that the apparent horizons are initially at rest; the tangential term
determines the black hole's spin~\cite{Cook2002,Cook2004,Caudill-etal:2006}. 

References~\cite{Cook2002,Cook2004,Caudill-etal:2006} chose
the sign of the 
last term in Eq.~(\ref{eq:InnerShiftBC}) such that positive values 
of $\Omega_r$ counteract the spin of the corotating holes that are obtained 
with $\Omega_r=0$. Here, we are interested in large spins, and we 
reverse the sign of the last term in Eq.~(\ref{eq:InnerShiftBC}) so that 
positive, increasing $\Omega_r$ results in increasing spins.

Two sets of choices for $\CMetric_{ij}$, $K$, $\BoundIn$, 
and the boundary 
condition for $\Lapse\CF$ on $\BoundIn$ are discussed in the next subsections. 
Each set of choices will be used to construct binary-black-hole initial 
data in Sec.~\ref{sec:BBHData}.
%%%%%%%%%%%%%%%%%%%%%%%%%%%%%%%%%%%%%%%%%%%%%%%%%%%%%%%%%%%%%%%%
\subsubsection{Conformal flatness \& maximal slicing (CFMS)}
\label{sec:FormalismCF}

The simplest choice for $\CMetric_{ij}$ is a flat metric,
\begin{equation}\label{eq:CMetricIsFlat}
\CMetric_{ij}\equiv\SFlatMetric_{ij}.
\end{equation}
This choice
has been used almost exclusively in the previous formulations of 
binary-black-hole initial data.

The simplest choice for $\ExCurv$, also commonly used in prior formulations of 
binary-black-hole initial data, is maximal slicing, i.e.
\begin{equation}\label{eq:SliceIsMaximal}
K\equiv 0.
\end{equation}

Also for simplicity, we choose to make the excision surface $\BoundIn$ 
consist of coordinate spheres:
\begin{equation}
\BoundIn = \bigcup_{a=1}^n \BoundIn_a,
\end{equation} where $\BoundIn_a$ are  
surfaces of constant Euclidean distance $\rexc$ about the center of 
each excised hole, and $n=1\mbox{ or }2$ is the number of black holes 
present in the initial data. 

The boundary condition for the lapse on 
$\BoundIn$ determines the temporal gauge;
we adopt the condition given in  Eq.~(59a) of Ref.~\cite{Cook2004}:
\begin{equation}\label{eq:VonNeumannLapseBC}
\frac{\partial}{\partial r_a} (\Lapse\CF) = 0 \mbox{ on } \BoundIn_a,
\end{equation} 
where $r_a$ is the Euclidean distance from the center of hole $a$.
This type of initial data is used in Refs.~\cite{Scheel2006,Pfeiffer-Brown-etal:2007,Scheel2008}.

%%%%%%%%%%%%%%%%%%%%%%%%%%%%%%%%%%%%%%%%%%%%%%%%%%%%%%%%%%%%%%%%

\subsubsection{Superposed Kerr Schild (SKS)}
\label{sec:FormalismSKS}
Single black holes with angular~\cite{GaratPrice:2000,Kroon:2004} or 
linear~\cite{york80} momentum 
do not admit conformally-flat spatial slicings;  therefore,
conformal flatness [Eq.~(\ref{eq:CMetricIsFlat})] is necessarily deficient. 
This has motivated 
investigations of binary-black-hole initial data whose free data 
have stronger physical motivation, e.g.
Refs.~\cite{Matzner1999,Marronetti-Matzner:2000,Pfeiffer2002a,Tichy2002,%
Nissanke2006,Yunes2006a,Yunes2006b,hannamEtAl:2007,kellyEtAl:2007}.

In this subsection, we consider conformally-curved data that 
are in the same 
spirit as the SKS data of Refs.~\cite{Matzner1999,Marronetti-Matzner:2000}
although here i) we apply the idea 
to the QE-XCTS formalism, and ii) as discussed below, our 
free data is very nearly conformally-flat and maximally-sliced 
\emph{everywhere except in the vicinity of the black holes}. 

%%%%%%%%%%%%%%%%%%%%%%%%%%%%%%%%%%%%%%%%%%%%%%%%%%%%%%%%%%%%%%%%
\begin{table*}
\begin{tabular}{|lcc | c c  c  c  c c | c c c c |}
\hline
Label  & Section & Figures
& n &   d  & $\Omega_0$ & $\dot{a}_0\times 10^4$ & $\Omega_r$ or $S/m_p^2$
&  $\tilde{S}$ 
& $\left|\TypeAKV{\SMM}\right|$ & $\Mirr$ & $M$ & $\Eadm$\\\hline
BY-Single & \ref{sec:SingleBHPunctureData}  
& \ref{fig:PunctureConvTest}--\ref{fig:PsiAngularStructure}, 
\ref{fig:PunctureXCTS_EmbedDiag}, \ref{fig:DeltaChi}
& 1 & - & - & - & $0.01 \leq S/m_p^2 \leq 10^4$ & - & & & & \\
CFMS-Single & \ref{sec:SingleBHQEXCTS} 
& \ref{fig:CFBHMassSpinUnscaled}--\ref{fig:PunctureXCTS_EmbedDiag}, 
\ref{fig:DeltaChi}
& 1 & - & - & - & $0 \leq \Omega_r \leq 0.191$ & - & & & &\\
CFMS   & \ref{sec:CFBBHID} 
& \ref{fig:BBH_CFMS_Dimensionless}, \ref{fig:TimeDerivatives}
& 2 & 32  & 0.007985 & 0         & $0\le\Omega_r\le0.1615$ & -    & 
  & & &\\\hline
SKS-0.0 & \ref{sec:SKSBBHID} 
&  \ref{fig:IDPlot_SKS_MassSpin},~\ref{fig:TimeDerivatives}
& 2 & 32  & 0.006787 & 0         & $0\le\Omega_r\le0.24$    & 0 &  
  & & & \\
SKS-0.5 & \ref{sec:SKSBBHID} 
&  \ref{fig:IDPlot_SKS_MassSpin},~\ref{fig:TimeDerivatives}
& 2 & 32  & 0.006787 & 0         & $0\le\Omega_r\le0.27$    & 0.5  & 
  & & &\\ 
SKS-0.93 & \ref{sec:SKSBBHID} 
&  \ref{fig:IDPlot_SKS_MassSpin}--\ref{fig:TimeDerivatives}
& 2 & 32  & 0.006787 & 0         & $0\le\Omega_r\le0.35$    & 0.93 & 
 & & & \\
SKS-0.99 & \ref{sec:SKSBBHID} 
&  \ref{fig:IDPlot_Convergence}--\ref{fig:TimeDerivatives}
& 2 & 32  & 0.007002 & 3.332         & $0.28\le\Omega_r\le0.39$    & 0.99 & 
 & & & \\\hline
SKS-0.93-E0  & \ref{sec:EccentricityReduction} 
&  \ref{evFig:dsdtVsTimeBothInMadmUnits}
& 2 & 32  & 0.006787    & 0         & 0.28   & 0.93 
  & 0.9278 & 0.9371  & 1.131 & 2.243\\
SKS-0.93-E1  & \ref{sec:EccentricityReduction} 
&  \ref{evFig:dsdtVsTimeBothInMadmUnits}
& 2 & 32  & 0.007    & 0         & 0.28   & 0.93 
  & 0.9284 & 0.9375  & 1.132 & 2.247\\
SKS-0.93-E2  & \ref{sec:EccentricityReduction} 
&  \ref{evFig:dsdtVsTimeBothInMadmUnits}
& 2 & 32  & 0.006977 & 3.084 & 0.28   & 0.93 
  & 0.9275 & 0.9395  & 1.134 & 2.249\\
SKS-0.93-E3   & \ref{sec:evCirc} 
&  \ref{fig:IDPlot_Convergence}--\ref{fig:IDPlot_SKS_MassSpin},
\ref{fig:TimeDerivatives}--\ref{evFig:SpinDiffAndDiffSpinPaper},
\ref{fig:DeltaChi}
& 2 & 32  & 0.007002 & 3.332 & 0.28  & 0.93 & 0.9275 & 0.9397 & 1.134 
& 2.250 \\
SKS-HeadOn   & \ref{sec:evPlunge} 
& \ref{fig:IDPlot_Convergence}--\ref{fig:IDPlot_SKS_MassSpin}, 
\ref{fig:TimeDerivs}, 
%\ref{fig:d100Convergence}--\ref{fig:d100SpinFromShape}
\ref{fig:d100Convergence}--\ref{fig:DeltaChi}
& 2 & 100 & 0      & 0    & 0.3418& 0.97 & 0.9701 & 0.8943 & 1.135 
& 2.257  \\\hline
\end{tabular}
\caption{Summary of the initial data sets constructed in this paper.
  The first row (BY-Single) represents Bowen-York
  initial data for single black holes of various spins.  The
  next two rows (CFMS-Single and CFMS) are quasi-equilibrium, conformally-flat,
  maximally-sliced initial data for single and binary spinning black
  holes, respectively.  All other data sets employ superposed
  Kerr-Schild quasi-equilibrium data with the second block of
    rows representing {\em families} of initial data sets for various
    spins and the last block of rows representing {\em  individual}
    data sets to be evolved.  The 
  data sets SKS-0.93-E0 to
  SKS-0.93-E3 demonstrate eccentricity removal, and 
  SKS-HeadOn is used in a head-on evolution.  The first block
  of columns gives the label used for each data set, and the relevant
  section of this paper devoted to it.  The next block of columns
  lists the most important parameters entering the initial data.  The
  last block of columns lists some properties of those data sets that
  we evolve in Sec.~\ref{sec:Evolutions}.
\label{tab:IDSummary}}
\end{table*}
%%%%%%%%%%%%%%%%%%%%%%%%%%%%%%%%%%%%%%%%%%%%%%%%%%%%%%%%%%%%%%%%

The choices we make here generalize the conformally-curved data 
in chapter 6 of Ref.~\cite{Thesis:Lovelace} to nonzero spins. 
Specifically, the free data and lapse boundary condition 
will be chosen so that the conformal geometry near each hole's horizon is that 
of a boosted, spinning, Kerr-Schild black hole. 
The conformal metric $\tilde{g}_{ij}$ and the mean curvature $K$
take the form
\begin{eqnarray}\label{eq:CMetricSKS}
\tilde{g}_{ij} & := & f_{ij}
+\sum_{a=1}^{n}e^{-r_a^2/w_a^2}\left(g_{ij}^a - f_{ij}\right),\\
K & := & \sum_{a=1}^{n} e^{-r_a^2/w_a^2}K_a.\label{eq:TrKSKS}
\end{eqnarray} Here $g_{ij}^{a}$ and $K_a$ are the spatial metric 
and mean curvature, respectively, of a boosted, 
spinning Kerr-Schild black
hole with mass $\tilde{M}_a$, spin $\tilde{S}_a$, and speed $\tilde{v}_a$.

Far from each hole's horizon, the conformal metric is very nearly flat; 
this prevents the 
conformal factor from diverging on the outer boundary~\cite{Thesis:Lovelace}.
The parameter $w_a$ is a weighting factor that determines how quickly the 
curved parts of the conformal data decay with Euclidean distance 
$r_a\mbox{ }(a=1,2,...)$ from hole $a$;  in this paper, 
the weight factor $w_a$ is chosen to be larger than 
the size scale of hole $a$ but smaller than the distance $d$ to the 
companion hole (if any): $M_a \lesssim w_a \lesssim d_a$. 
This is similar to the ``attenuated'' superposed-Kerr-Schild data of 
Refs.~\cite{Marronetti-Matzner:2000,BonningEtAl:2003}, 
except that here the weighting functions are Gaussians which 
vanish far from the holes, while in 
Refs.~\cite{Marronetti-Matzner:2000,BonningEtAl:2003} the weighting functions 
go to unity far from the holes.

The excision surfaces $\BoundIn_a$ are not coordinate 
spheres unless $\tilde{S}_a=0$ and $\tilde{v}_a = 0$. Instead
they are deformed in two ways. i) They are distorted so that 
 they are surfaces of 
constant Kerr radius $r_{\rm Kerr}$, i.e.
\begin{equation}
\frac{x^2+y^2}{r_{\rm Kerr}^2+\tilde{S_a}^2/\tilde{M_a}^2}
+\frac{z^2}{r_{\rm Kerr}^2}=1
\end{equation}
where $x$, $y$, and $z$ are Cartesian 
coordinates on the $\BoundIn$.  Then,
ii) the excision surfaces are Lorentz-contracted along the direction of 
the boost.

The boundary condition for the lapse $\Lapse$ 
on $\BoundIn_a$ is a Dirichlet condition that 
causes $\Lapse$ (and, consequently, the temporal gauge) in the 
vicinity of each hole to be nearly that of the corresponding Kerr-Schild 
spacetime, i.e.
\begin{eqnarray}\label{eq:SKSLapseBC}
\alpha\psi = 1 + \sum_{a=1}^n e^{-r_a^2/w_a^2} (\alpha_a-1) 
\mbox{ on }\BoundIn_a,
\end{eqnarray} where $\alpha_a$ is the lapse corresponding to the Kerr-Schild 
spacetime $a$.

%%%%%%%%%%%%%%%%%%%%%%%%%%%%%%%%%%%%%%%%%%%%%%%%%%%%%%%%%%%%%%%%
\section{Single-black-hole initial data with nearly-extremal spins}
\label{sec:SBHdata}\label{sec:SBHData}
%%%%%%%%%%%%%%%%%%%%%%%%%%%%%%%%%%%%%%%%%%%%%%%%%%%%%%%%%%%%%%%%

In this section, we examine to which extent the formalisms presented
in Sec.~\ref{sec:IDF} can generate single black hole initial data with
nearly-extremal spin.  We consider first Bowen-York initial data and then 
conformally-flat quasi-equilibrium data.  Since superposed-Kerr-Schild data 
can represent single Kerr black holes exactly, there is no need to
investigate single-hole superposed-Kerr-Schild data. 
In Sec.~\ref{sec:BBHData}, we will both consider conformally-flat and 
superposed-Kerr-Schild data for binary black holes. 

To orient the reader, the initial data sets constructed in this section, 
as well as the binary-black-hole data sets constructed in 
Sec.~\ref{sec:BBHData}, are summarized in Table~\ref{tab:IDSummary}.

Unless noted otherwise, all spins presented in this section
are measured using the approximate-Killing-vector spin $\AKVSMM$
described in Appendix~\ref{sec:QuasilocalSpin}.  Therefore, the
subscript ``AKV'' in $\AKVSMM$ will be suppressed for simplicity.

\subsection{Bowen-York (puncture) initial data}
\label{sec:SingleBHPunctureData}

As discussed in Sec.~\ref{sec:PunctureData}, for a single spinning
black hole at rest, puncture initial data is {\em identical} to 
inversion-symmetric initial data.  Such solutions have been examined in the
past (e.g.~\cite{yorkpiran,cook90}), and additional results were
obtained (partly in parallel to this work) in the study by Dain, Lousto,
 and Zlochower {\em al}~\cite{DainEtAl:2008}.

We revisit this topic here 
to determine the
maximum possible spin of Bowen-York (BY) initial data more accurately than
before, to establish the power-law coefficients for the approach to these
limits with increasing spin parameter $S$, and to present new results
about the geometric structure of Bowen-York initial data with very
large spin parameter.

%%%%%%%%%%%%%%%%%%%%%%%%%%%%%%%%%%%%%%%%%%%%%%%%%%%%%%%%%%%%%%%%
\begin{figure}
\includegraphics[scale=0.52]{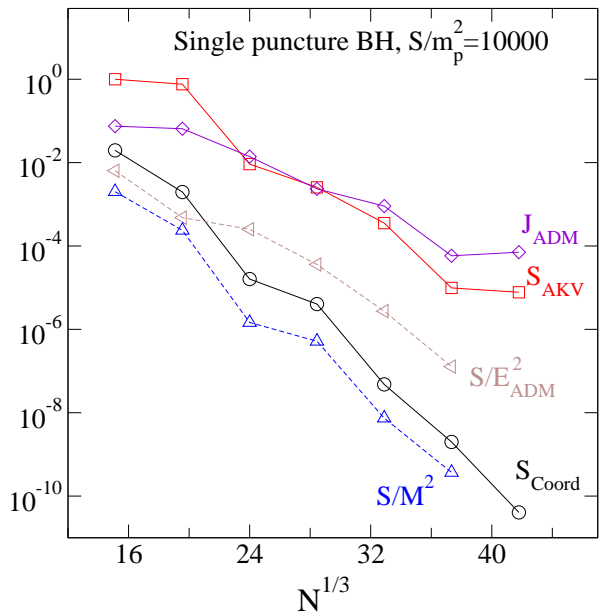}
\caption{ \label{fig:PunctureConvTest} Convergence test for a single
  puncture black hole with a very large spin parameter 
  $S/m_p^2=10000$. 
  Plotted are results vs. resolution $N$, which is 
  the total number of basis-functions.  
  The solid lines show the relative differences of
  three angular momentum measures to the analytically expected value
  $10000$.  The dashed lines show differences 
  from the next-higher
  resolution of two dimensionless quantities
  for which no analytic answer is available. }
\end{figure}
%%%%%%%%%%%%%%%%%%%%%%%%%%%%%%%%%%%%%%%%%%%%%%%%%%%%%%%%%%%%%%%%

We solve Eq.~(\ref{eq:Puncture-u-Eqn}) with the pseudo-spectral elliptic solver
described in Ref.~\cite{Pfeiffer2003}.  The singular point of $u$ at
the origin is covered by a small rectangular block extending from $\pm
10^{-4}m_p$ along each coordinate axis.  This block overlaps four
concentric spherical shells with radii of the boundaries at $8\cdot
10^{-5}m_p, 0.005m_p, 0.3m_p, 50m_p,$ and $10^9m_p$.  The equations
are solved at several different resolutions, with the highest
resolution using $20^3$ basis-functions in the cube, $L=18$ in the
spheres and 26 and 19 radial basis-functions in the inner and outer
two spherical shells, respectively.  

Because of the axisymmetry of the data-set, the rotational Killing vector
of the apparent horizon is simply $\partial_\phi$.  The integral for
the quasilocal spin, Eq.~(\ref{eq:Spin}) turns out to be independent
of $\CF$ and can be evaluated analytically with a result equal to the
spin-parameter, $S$.  Thus we can use this initial data set to check
how well our spin-diagnostics and our ADM angular momentum diagnostic
works (recall that $\Jadm$ is also equal to the spin-parameter $S$).
This comparison is performed in Fig.~\ref{fig:PunctureConvTest}, which
shows relative differences between the numerically extracted values
for the approximate-Killing-vector (AKV) spin, 
the coordinate spin (defined with the AKV spin 
in Appendix~\ref{sec:QuasilocalSpin}), 
and the ADM angular momentum $\Jadm$ relative to the expected
answer, $S$.  The figure also shows differences between
neighboring resolutions for the two quantities of interest below,
$S/M^2=\SMM$ and $S/\Eadm^2=\Jadm/\Eadm^2=\JEE$.

%%%%%%%%%%%%%%%%%%%%%%%%%%%%%%%%%%%%%%%%%%%%%%%%%%%%%%%%%%%%%%%%
\begin{figure}
\includegraphics[scale=0.52]{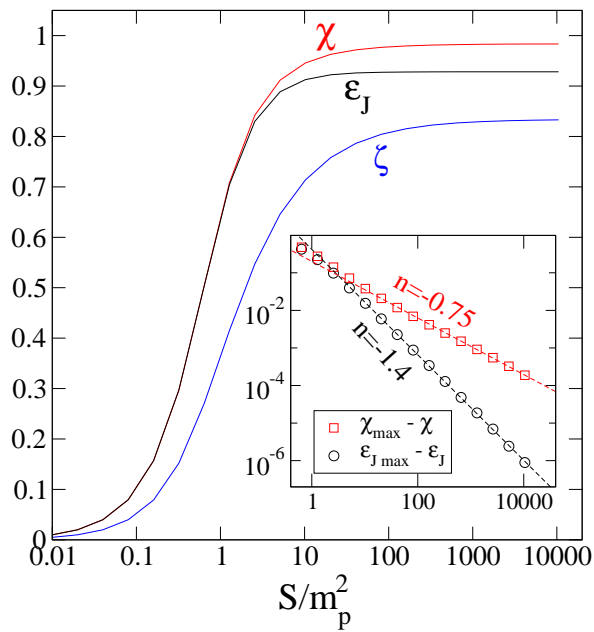}
\caption{\label{fig:puncture} Properties of single, spinning
  puncture black holes with spin-parameter $S$ and puncture mass
  $m_p$.  The
  dimensionless spin $\SMM:=S/M^2$,
  ADM angular momentum $\varepsilon_J:= J_{\rm ADM}/E_{\rm ADM}^2$, 
  and spin-extremality parameter $\SoTwoMirrMirr:=S/\left(2\Mirr^2\right)$
  are plotted against the spin parameter $S/m_p^2$. 
  The horizon mass $M$ is related to the spin $S$ and 
  irreducible mass $\Mirr$ in Eq.~(\ref{eq:GoodSpinNorm}).}
\end{figure}
%%%%%%%%%%%%%%%%%%%%%%%%%%%%%%%%%%%%%%%%%%%%%%%%%%%%%%%%%%%%%%%%

Figure~\ref{fig:PunctureConvTest} seems to show
  exponential convergence with increased resolution $N$.  Since
  puncture data is only $C^2$ at the puncture, one would rather expect
  polynomial convergence.  The effect of the non-smoothness at the
  puncture is mitigated by choosing a very high resolution close to
  the puncture (a small cube with sides $\pm 10^{-4}m_p$ with $20^3$
  basis-functions).  Therefore, for the resolutions considered in
  Fig.~\ref{fig:PunctureConvTest}, the truncation error is dominated
  by the solution away from the puncture, and exponential
  convergence is visible.  If we used infinite-precision arithmetic
  and were pushing toward higher resolution than shown in 
  Fig.~\ref{fig:PunctureConvTest}, then we would expect to eventually see
  polynomial convergence dominated by the cube covering the
  puncture.

Next, we construct a series of initial data sets with increasing
spin-parameter $S$, and compute $\SMM$, $\JEE$, and $\SoTwoMirrMirr$ 
for each initial data set.
The results are plotted in Fig.~\ref{fig:puncture} and confirm
earlier results~\cite{cook90,DainEtAl:2002}.  In addition, the
inset shows that the asymptotic values $\SMM_{\rm max}=0.9837$ and
$\varepsilon_{u,\rm max}=0.928200$ are approached as {\em power-laws}
in the spin-parameter,
\begin{align}\label{eq:puncture-powerlaw1}
\SMM_{\rm max}-\SMM &\propto \left(\frac{S}{m_p^2}\right)^{-0.75},\\
\label{eq:puncture-powerlaw2}
\varepsilon_{J,\rm max}-\JEE &\propto \left(\frac{S}{m_p^2}\right)^{-1.4}.
\end{align}
The exponents of these power-laws are computed here for the first time.

\begin{figure}
\includegraphics[scale=0.52]{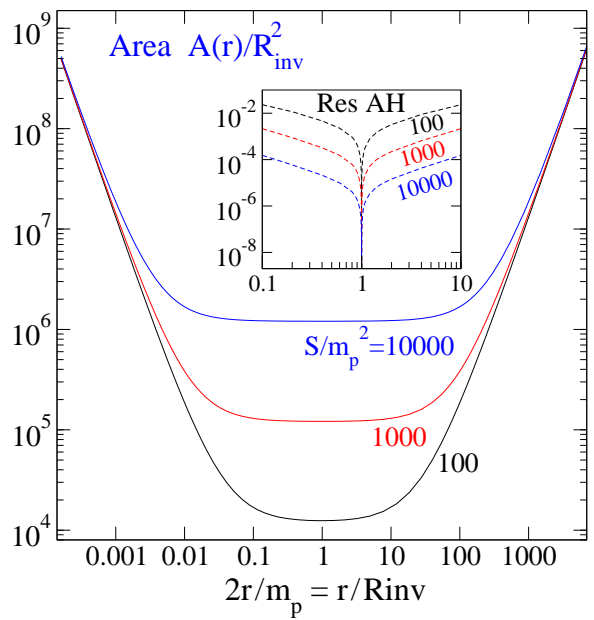}
\caption{\label{fig:PunctureArea}
  Properties of coordinate spheres with radius $r$ for 
  high-spin puncture initial data.  
  \emph{Main panel:} Area
  of these spheres. \emph{Inset:} residual of the apparent horizon equation on
  these spheres.  The area is almost constant over several orders of magnitude
  in $r$.  The apparent-horizon-residual 
  vanishes at $r=R_{\rm inv}$, but is very
  small over a wide range of $r$.}
\end{figure}

To confirm that the apparent horizon is indeed at $r=R_{\rm
  inv}$, we ran our apparent horizon finder on the high-spin
puncture initial data sets.  The horizon finder had great difficulty  
converging, and the reason for this becomes clear from
Fig.~\ref{fig:PunctureArea}. The main panel of this figure shows the 
area of spheres with
coordinate radius $r$.  The area is minimal at $r=m_p/2$, as it must be,
since $m_p/2=R_{\rm inv}$ is the radius of the inversion sphere.
However, the area is almost constant over a wide range in $r$---for
$S/m_p^2=10000$ over about two decades in either 
direction: $0.01\lesssim
r/R_{\rm inv}\lesssim 100$.  Thus, the Einstein-Rosen bridge (the
throat) connecting the two asymptotically flat universes lengthens as
the spin increases, giving rise to an ever-lengthening cylinder.  If
this were a perfect cylinder, then the expansion would be zero for any
$r=\mbox{const}$ cross-section.  Because the geometry is not perfectly
cylindrical, the expansion vanishes only for $r=m_p/2=R_{\rm inv}$,
but remains very small even a significant distance away from
$r=m_p/2=R_{\rm inv}$.  This is shown in the inset, which plots the
residual of the apparent horizon finder at different radii.

With the
lengthening of the throat, the interval in $r$ with small expansion
lengthens, and the value of the expansion within this interval
reduces. Both effects make it harder for the apparent horizon finder to
converge.  In Fig.~\ref{fig:puncture}, we have used our knowledge of
the location of the apparent horizon to {\em set} $r_{\rm AH}=m_p/2$,
rather than to find this surface numerically.  Without this knowledge,
which arises due to the identification of puncture data and inversion
symmetric data, computation of Fig.~\ref{fig:puncture} would have been
significantly harder, perhaps impossible.  

Let us assume for the moment that the solution
$\CF(r)=\frac{m_p}{2r}+1+u(r)$ is spherically symmetric (we give
numerical evidence below that this is indeed a good approximation).
Because $\SMetric_{ij}=\CF^4f_{ij}$, the area of coordinate spheres is
then given by
\begin{equation}
A(r) = 4\pi \CF^2(r) r.
\end{equation}
In the throat region, where $A(r)\approx\mbox{const}$,
the  conformal factor must therefore behave like $1/\sqrt{r}$, as
also argued independently by Dain, Lousto, and Zlochower~\cite{DainEtAl:2008}.

To extend on Dain et {\em al.}'s analysis, let us substitute
Eq.~(\ref{eq:A-BY-S}) into Eq.~(\ref{eq:Ham3}) to obtain the
well-known equation
\begin{equation}\label{eq:puncture-for-spin}
\CCD^2\CF=-\frac{9S^2\sin^2\theta}{4r^6}\CF^{-7},
\end{equation}
where $\theta$ is the angle between the spin-direction and the point $x^i$.  
Continuing to assume that $\CF$ is
approximately spherically symmetric, we can replace the factor
$\sin^2\theta$ by its angular average
$(4\pi)^{-1}\int \sin^2\theta\,d\Omega = 2/3$, and obtain
\begin{equation}\label{eq:SphericalSymmetricPuncture}
\frac{d^2\bar\CF}{dr^2} + \frac{2}{r}\frac{d\bar\CF}{dr}
=-\frac{3S^2}{2r^6}\bar\CF^{-7}.
\end{equation}
Here, we introduced an overbar $\bar\CF$ to distinguish the
spherically symmetric solution $\bar\CF(r)$ of
Eq.~(\ref{eq:SphericalSymmetricPuncture}) from the full solution
$\CF(x^i)$ of puncture/inversion-symmetric initial data.  Following
Dain et {\em al.}~\cite{DainEtAl:2008} we assume that the conformal
factor behaves as a power-law ($\bar\CF(r)=A r^\alpha$) and substitute
this into Eq.~(\ref{eq:SphericalSymmetricPuncture}).  We find that
Eq.~(\ref{eq:SphericalSymmetricPuncture}) determines the power-law
exponent $\alpha=-1/2$ {\em and} the overall amplitude
$A=(6S^2)^{1/8}$, so that

\begin{equation}\label{eq:Sqrt-r}
\bar\CF(r)=\frac{\left(6S^2\right)^{1/8}}{\sqrt{r}}
%=\frac{\left[6 \left(S/m_p^2\right)^2\right]^{1/8}}{\sqrt{r/m_p}}
=96^{1/8}\left(\frac{S}{m_p^2}\right)^{1/4}\left(\frac{r}{R_{\rm inv}}\right)^{-1/2}.
\end{equation}
In Eq.~(\ref{eq:Sqrt-r}), we
chose the scaling $S/m_p^2$ which is commonly used in the
puncture-data literature, but kept $r/R_{\rm inv}$ to emphasize the
inversion symmetry of the data in our figures (in a log-plot using
$r/R_{\rm inv}$, the solution will appear symmetric, see e.g. 
Fig.~\ref{fig:PunctureArea}).  While $\bar\CF(r)$  solves 
the spherically symmetric
Eq.~(\ref{eq:SphericalSymmetricPuncture}) exactly, it must deviate
from $\CF(x^i)$ for sufficiently large $r$ because $\bar\CF\to 0$ as
$r\to\infty$, whereas $\CF\to 1$.  
The deviation will become significant when
$\bar\CF\sim 1$, i.e. at radius $r_x\sim \sqrt{S/m_p^2}$.  Because of
inversion symmetry, this implies a lower bound of validity at $1/r_x$, 
so that Eq.~(\ref{eq:Sqrt-r}) holds for 
\begin{equation}\label{eq:Sqrt-r-range}
\left(\frac{S}{m_p^2}\right)^{-1/2}
\lesssim \frac{r}{R_{\rm inv}}\lesssim 
%\left(\frac{3}{2048}\right)^{1/4} 
\left(\frac{S}{m_p^2}\right)^{1/2}.
\end{equation}

The circumference of the cylindrical throat is 
\begin{equation}\label{eq:PunctureThroat-LC} 
{\cal C}=2\pi \bar\CF(r)^2 r=2\pi 96^{1/4}\sqrt{\frac{S}{m_p^2}} R_{\rm inv},
\end{equation}
and its length is
\begin{equation}\label{eq:PunctureThroat-L} 
{\cal L}=\int_{(S/m_p^2)^{-1/2}}^{(S/m_p^2)^{1/2}}\bar\CF^2(r)\, dr=96^{1/4}\sqrt{\frac{S}{m_p^2}}\ln\left(\frac{S}{m_p^2}\right)\,R_{\rm inv}.
\end{equation}
Therefore, the ratio of length to circumference, 
\begin{equation}\label{eq:PunctureThroat-LC-Ratio}
\frac{\cal L}{\cal C}=\frac{1}{2\pi}\ln\left(\frac{S}{m_p^2}\right),
\end{equation}
grows without bound as $S/m_p^2$ becomes large, albeit very slowly.
The scaling with $(S/m_p^2)^{1/2}$ in
  Eqs.~(\ref{eq:Sqrt-r-range})--(\ref{eq:PunctureThroat-L}) might seem
  somewhat surprising.  However, in the large spin limit, $S/M^2$ is
  just a constant close to unity (namely $\chi_{\rm max}=0.9837$).
  Therefore, $S^{1/2}\approx M$, i.e. the scaling $S^{1/2}$ is
  effectively merely a scaling with mass.

%%%%%%%%%%%%%%%%%%%%%%%%%%%%%%%%%%%%%%%%%%%%%%%%%%%%%%%%%%%%%%%%
\begin{figure}
\includegraphics[scale=0.52]{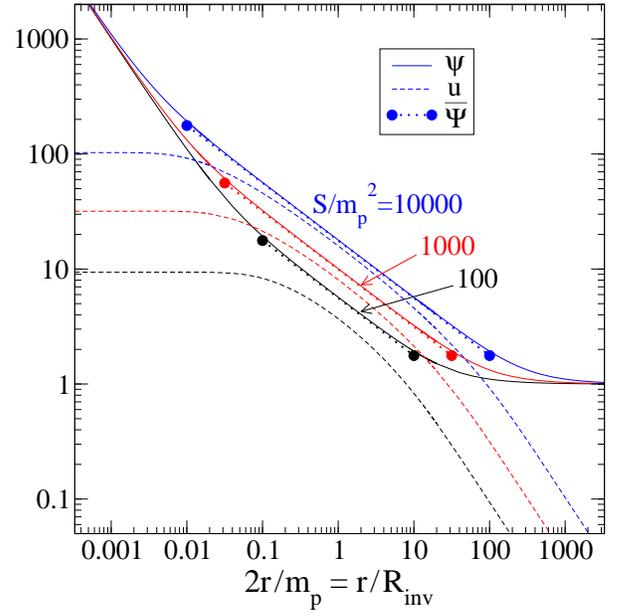}
\caption{\label{fig:Psi_u}Solutions of high-spin puncture
  initial data.  Plotted 
  are the conformal factor $\CF$ and 
  puncture function $u$ 
  in the equatorial plane as a
  function of radius $r$.  
  Furthermore, the approximate solution $\bar\CF$
  is included, with solid circles denoting the range of validity of
  this approximation, cf. Eq.~(\ref{eq:Sqrt-r-range}).  Three curves
  each are plotted, corresponding from top to bottom to
  $S/m_p^2=10000, 1000, 100$.}
\end{figure}
%%%%%%%%%%%%%%%%%%%%%%%%%%%%%%%%%%%%%%%%%%%%%%%%%%%%%%%%%%%%%%%%

Figure~\ref{fig:Psi_u} shows the conformal factor $\CF$, the
``puncture function'' $u$, and the estimate $\bar\CF$ of
Eq.~(\ref{eq:Sqrt-r}) for three different values of $S/m_p^2$.  There
are several noteworthy features in this figure.  First, both $\CF$ and
$u$ show clearly three different regimes: 
\begin{itemize}
\item For large $r$, $\CF\approx 1$ and $u\propto 1/r$.  This is the
  upper asymptotically-flat end.  
\item For intermediate $r$, $\CF\propto 1/\sqrt{r}$ and $u\propto
  1/\sqrt{r}$.  This is the cylindrical geometry extending
  symmetrically around the throat.
  This region becomes more pronounced as $S$ increases.
\item For small $r$, $\CF\propto 1/r$ and $u\approx\mbox{const}$.
  This is the lower asymptotically-flat end.
\end{itemize}
Figure~\ref{fig:Psi_u} also plots the approximate
solution $\bar\CF$ [cf. Eq.~(\ref{eq:Sqrt-r})] for its range of
validity [given by Eq.~(\ref{eq:Sqrt-r-range})].  Note that slope {\em
  and} amplitude of $\bar\CF$ fit very well the numerical
solution $\CF$.  In fact, the agreement is much better than with $u$.

One could also have started the calculation that led to
Eq.~(\ref{eq:Sqrt-r}) with Eq.~(\ref{eq:Puncture-u-Eqn}).  Assuming
spherical symmetry, {\em and} assuming that $u\gg m_p/(2r)+1$, we
would have derived Eq.~(\ref{eq:SphericalSymmetricPuncture}), but with
{\em $\bar\CF$ replaced by $u$}.  We would then have found the
approximate behavior Eq.~(\ref{eq:Sqrt-r}) {\em for $u$}.  The
disadvantage of this approach is the need for additional
approximations, which reduce the accuracy of the result.  From
Fig.~\ref{fig:Psi_u} we see that, in the throat region, the dotted
lines representing $\bar\CF$ are close to the dashed lines of $u$.
But the agreement between $\CF$ and $\bar\CF$ is certainly better.

Finally, we note that the limits of validity of $\bar\CF$
[Eq.~(\ref{eq:Sqrt-r-range})] match very nicely the points where the
numerical $\CF$ diverges from $\bar\CF$.

%%%%%%%%%%%%%%%%%%%%%%%%%%%%%%%%%%%%%%%%%%%%%%%%%%%%%%%%%%%%%%%%
\begin{figure}
\includegraphics[scale=0.52]{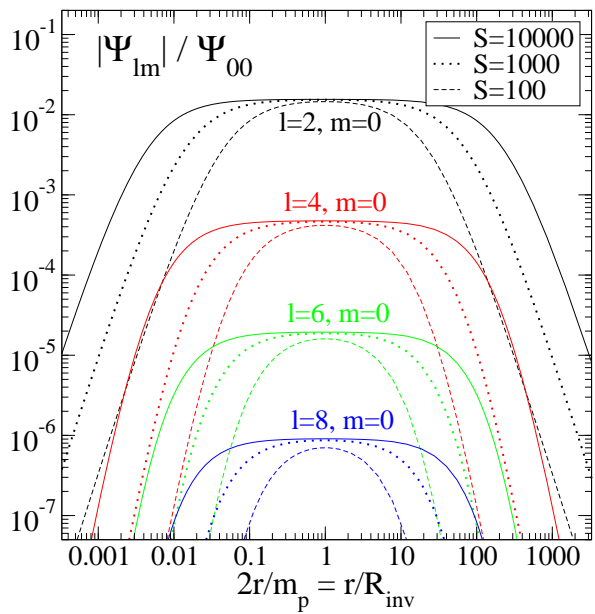}
\caption{\label{fig:PsiAngularStructure} Angular decomposition of the
  conformal factor $\CF(r, \theta,\phi)$ for single black hole
  puncture data.}
\end{figure}
%%%%%%%%%%%%%%%%%%%%%%%%%%%%%%%%%%%%%%%%%%%%%%%%%%%%%%%%%%%%%%%%

To close this section, we present numerical evidence that indeed $\CF$
is approximately spherically symmetric, the assumption that entered
into our derivation of Eq.~(\ref{eq:Sqrt-r}). We decompose the
conformal factor of the numerical puncture data solutions into
spherical harmonics,
\begin{equation}
\CF(r,\theta,\phi)=\sum_{l=0}^{\infty}\sum_{m=-l}^{l}
\CF_{lm}(r)Y_{lm}(\theta,\phi),
\end{equation}
and plot in Fig.~\ref{fig:PsiAngularStructure} the sizes of the $l\neq
0$ modes relative to the spherically symmetric mode
$\CF_{00}$. Because of the symmetries of the problem, 
the only non-zero modes have $m=0$ and
even $l$.  In the throat region, the largest non-spherically symmetric
mode $\CF_{20}$ is about a factor of 65 smaller than the spherically
symmetric mode.  With increasing $l$, $\CF_{lm}$ decays very rapidly.
Also, in both asymptotically flat ends, the non-spherically symmetric
modes decay more rapidly than the $l=0$ mode, as expected for
asymptotically flat data.  This figure again shows nicely the
inversion symmetry of the data, under $r/R_{\rm inv} \to (r/R_{\rm
  inv})^{-1}$.  Given the simple structure of the higher modes, it
should be possible to extend the analytical analysis of the throat to
include the non-spherical contributions.  To do so, one would expand
$\CF$ as a series in Legendre polynomials in $\theta$; 
the $\CF^{-7}$-term on the
right hand side of Eq.~(\ref{eq:puncture-for-spin}) would result in a
set of ordinary differential equations for those coefficients.  In the
throat region, the radial behavior of each mode should be $\propto
1/\sqrt{r}$, and the ordinary differential equations should simplify
to algebraic relations.

%#######################################
\subsection{Quasi-equilibrium extended-conformal-thin-sandwich data}
%#######################################
\label{sec:SingleBHQEXCTS}
We have seen in Sec.~\ref{sec:SingleBHPunctureData} that puncture initial data 
for single, spinning black holes can be constructed for holes with initial 
spins of $\SMM \le 0.9837$. In this section, we address the analogous 
question for excision black-hole initial data: 
how rapid can the initial spin be for a single, spinning black hole 
constructed using quasiequilibrium, extended-conformal-thin-sandwich (QE-XCTS) 
initial data?

As noted previously, 
if the free data $\CMetric_{ij}$ and 
$\TrExCurv$ are chosen to agree 
with the analytic values for a Kerr black hole, 
$\SMetric_{ij}^{\rm Kerr}$ and $\TrExCurv^{\rm Kerr}$, then 
the QE-XCTS initial data can exactly represent a single
Kerr black hole. In this case, $\SMM=1$ is obtained trivially by choosing 
$\CSpin=\CMass^2=1$, where $\CMass$ and $\CSpin$ are the mass and spin, 
respectively, of the Kerr black hole described by the conformal 
metric.

%%%%%%%%%%%%%%%%%%%%%%%%%%%%%%%%%%%%%%%%%%%%%%%%%%%%%%%%%%%%%%%%
\begin{figure}
\includegraphics[scale=0.5]{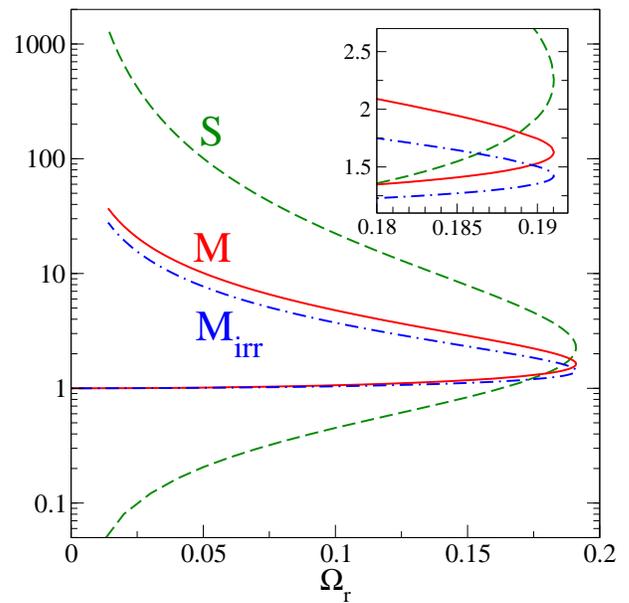}
\caption{
  Conformally-flat, maximally-sliced, quasiequilibrium
  initial data sets with a
  single, spinning black hole. We plot the horizon mass
  $\HorizonMass$, irreducible mass $\Mirr$, and the
  (approximate-Killing-vector) spin $\Spin$
  against the rotation parameter $\Omega_r$
  [cf. Eq.~(\ref{eq:InnerShiftBC})].
Only $\Omega_r$ is
  varied in this figure; all other parameters
  are held fixed. The
  upper and lower points with the same $\Omega_r$
  are obtained numerically by choosing different initial guesses.
  The inset shows a close-up view of the turning point, which 
  occurs at $\Omega_r\approx0.191.$
  \label{fig:CFBHMassSpinUnscaled}}
\end{figure}
%%%%%%%%%%%%%%%%%%%%%%%%%%%%%%%%%%%%%%%%%%%%%%%%%%%%%%%%%%%%%%%%

Setting aside this trivial solution, we construct 
conformally-flat, maximally-sliced (CFMS) data for a single, spinning hole.
We construct a family of QE-XCTS initial data sets for single 
spinning black holes by numerically solving the XCTS equations 
[in the form stated in Eqs.~(\ref{junkeq:XCTSa})--(\ref{junkeq:XCTSc})] 
using the same spectral elliptic solver~\cite{Pfeiffer2003} 
as in Sec.~\ref{sec:SingleBHPunctureData}.
The free data are given by 
Eqs.~(\ref{eq:CMetricIsFlat})--(\ref{eq:SliceIsMaximal}) and by 
Eqs.~(\ref{eq:freeTimeDerivsVanisha})--(\ref{eq:freeTimeDerivsVanishb}).

On the outer boundary $\BoundOut$, we impose 
Eqs.~(\ref{eq:OutBCStart})--(\ref{eq:OutBCEnd}). 
So that the coordinates are asymptotically inertial, 
we choose $\Omega_0=\dot{a}_0=0$ in Eq.~(\ref{eq:OutBCEnd}).

We excise a coordinate sphere of radius $\rexc$ about the origin, where 
\begin{equation}\label{eq:rexc-XCTS-CF}
\rexc=0.85949977
\end{equation}
is chosen 
such that for zero spin $\HorizonMass=1$. On this 
inner boundary $\BoundIn$, 
we impose Eqs.~(\ref{eq:AH-BC})--(\ref{eq:InnerShiftBC}) 
and Eq.~(\ref{eq:VonNeumannLapseBC}). The spin is determined by 
Eq.~(\ref{eq:InnerShiftBC}): first, the vector $\xi^i$ is chosen to 
be the coordinate rotation vector $\partial_\phi$, making the spin
point along the positive z axis; then,
the rotation parameter $\Omega_r$ 
is varied
while the other parameters are 
held fixed. The spin is measured on the apparent horizon using 
the approximate-Killing-vector spin 
(Appendix~\ref{sec:QuasilocalSpin});
because in this case the space is axisymmetric,
the ``approximate'' Killing vector reduces to the corresponding 
exact rotational Killing vector.

Figure~\ref{fig:CFBHMassSpinUnscaled} show how the mass $\HorizonMass$
and AKV spin $\Spin$ depend on $\Omega_r$.  At $\Omega_r=0$, we find
the spherically-symmetric solution with $S=0$ and $\Mirr=M=1$ (the
mass is proportional to the excision radius, and
Eq.~(\ref{eq:rexc-XCTS-CF}) sets it to unity).  Using this
spherically-symmetric solution as an initial guess for the elliptic
solver, we find solutions for increasing $\Omega_r$ with spin
increasing initially linearly with $\Omega_r$ and with approximately
constant mass.  Beyond some critical $\Omega_{\rm r, crit}$, the
elliptic solver fails to converge, and close to this point, all
quantities vary in proportion to $\sqrt{\Omega_{r,{\rm crit}}-\Omega_r}$.
These symptoms indicate a critical point where the solutions ``turn
over'' and continue towards smaller $\Omega_r$.  Analogous non-unique
solutions of the XCTS equations have been discovered before in
Ref.~\cite{Pfeiffer-York:2005}.  To construct solutions along the
upper branch, one must choose a sufficiently
close initial guess for the elliptic solver; we follow the steps
outlined in Ref.~\cite{Pfeiffer-York:2005} and are able to find
solutions along the upper branch for a wide range of
$\Omega_r<\Omega_{\rm r, crit}$.  As
Fig.~\ref{fig:CFBHMassSpinUnscaled} shows, mass and spin of the
horizon in solutions along the upper branch increase with decreasing
$\Omega_r$, analogous to the findings
in~\cite{Pfeiffer-York:2005,Baumgarte2007}.

%%%%%%%%%%%%%%%%%%%%%%%%%%%%%%%%%%%%%%%%%%%%%%%%%%%%%%%%%%%%%%%%
\begin{figure}
\includegraphics[scale=0.5]{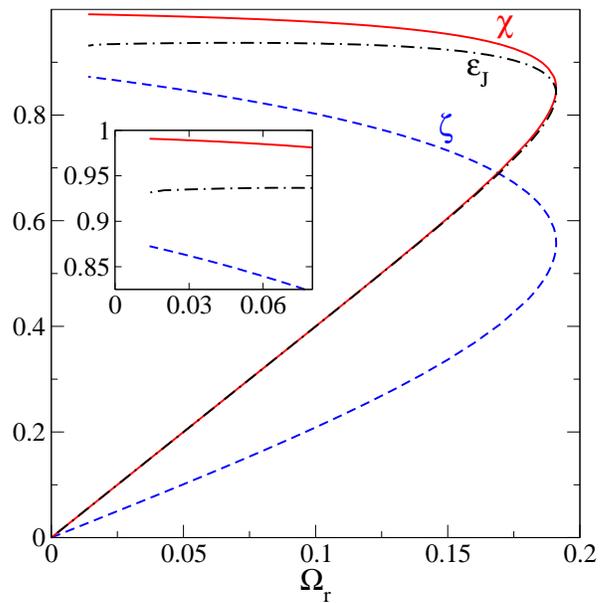}
\caption{
Conformally-flat, maximally-sliced
quasiequilibrium initial data sets with a single spinning black hole:
The dimensionless spin
$\SMM$, dimensionless ADM angular momentum $\JEE$, 
and spin-extremality parameter $\SoTwoMirrMirr$ plotted against
$\Omega_r$ [cf. Eq.~(\ref{eq:InnerShiftBC})].
Only $\Omega_r$ is
varied in this figure; all other parameters
are held fixed.
The inset enlarges the area in the upper left corner;
we are able to generate data sets with
$\SMM>0.99$, whereas the
largest spin obtainable on the lower branch is
$\SMM\approx 0.85$. 
\label{fig:CFBHMassSpin}}
\end{figure}
%%%%%%%%%%%%%%%%%%%%%%%%%%%%%%%%%%%%%%%%%%%%%%%%%%%%%%%%%%%%%%%%

Figure~\ref{fig:CFBHMassSpin} shows 
the dependence of $\SMM=\Spin/\HorizonMass^2$, 
$\JEE=\Jadm/\Eadm^2$, and $\SoTwoMirrMirr=S/\left(2\Mirr^2\right)$ 
on $\Omega_r$. 
The curves reflect again the non-unique solutions. 
The dimensionless spin $\SMM$ increases continuously along the
  lower branch, and reaches $\SMM\approx 0.85$ at the critical point.
  As $\Omega_r$ is decreased along the upper branch, $\SMM$ continues
  to increase, eventually reaching
  values larger than $0.99$.  It
  appears $\SMM$ continues to increase as $\Omega_r\to 0$.  To find
  the limiting value, consider that the behavior of the extremality
  parameter $\SoTwoMirrMirr$ in the inset of
  Fig.~\ref{fig:CFBHMassSpin}.  Assuming that $\SoTwoMirrMirr$ can be
  extrapolated to $\Omega_r\to 0$, we find a limiting value of
  $\SoTwoMirrMirr\approx 0.88$.  By Eq.~(\ref{eq:qFormula}), this 
implies a maximal value of $\SMM\approx 0.992$.  

In Figs.~\ref{fig:CFBHMassSpinUnscaled}--\ref{fig:CFBHMassSpin}, the
data sets on the lower branch appear to be physically reasonable.  For
spins $\SMM \lesssim 0.85$, the mass $\HorizonMass$ is nearly
constant, and the dimensionless spin $\SMM$ increases linearly with
$\Omega_r$.  Furthermore, as $\Omega_r\to 0$ the lower branch
  continuously approaches the exact Schwarzschild spacetime
  (see~\cite{Cook2004}).
The upper branch appears to be physically less reasonable; for instance, 
the spin $\SMM$ increases for {\em decreasing} horizon frequency $\Omega_r$. 
Comparing Figs.~\ref{fig:puncture} and~\ref{fig:CFBHMassSpin}, we see that the QE-XCTS data leads to somewhat 
larger values of $\SMM$ and $\varepsilon_J$ relative to puncture data. 
However, the values are not too different, and similar trends remain. For 
instance, $\SMM$ is much closer to unity than $\varepsilon_J$.  

To investigate
differences or similarities between puncture data and QE-XCTS data further, 
we compute embedding diagrams of the equatorial planes of these data sets.
The initial data for single black holes have rotational symmetry about the 
z-axis, so the metric (\ref{eq:CMetric}) on the initial data hypersurface, 
when restricted to the equatorial plane, can be written as
\begin{equation}\label{eq:CFBH_IDMetric}
ds^2
%\SMetric_{ij}
=\CF^4\left(dr^2+r^2d\phi^2\right),
\end{equation}
where $r$ and $\phi$ are the usual polar coordinates.   
This metric is now required to equal the induced metric
on the 2-D surface given by $\text{Z}=\text{Z}(\text{R})$ 
embedded in a 3-D Euclidean space with
line-element
\begin{equation}
ds^2_{\rm Euclidean}=d\text{R}^2 + \text{R}^2d\phi^2+d\text{Z}^2.
\end{equation}
Setting $d\text{Z}=\frac{d\text{Z}}{d\text{R}} d\text{R}$,
 we obtain the induced metric on the $\text{Z}=\text{Z}(\text{R})$ surface 

\begin{equation}\label{eq:EmbedMetric}
ds^2=
\left[1+\left(\frac{d\text{Z}}{d\text{R}}\right)^2\right]d\text{R}^2+\text{R}^2d\phi^2.
\end{equation}
Equating Eqs.~(\ref{eq:CFBH_IDMetric}) and~(\ref{eq:EmbedMetric}),
we find
\begin{equation}
\label{eq:Embed_R_vs_r}
\text{R} =\CF^2r
\end{equation}
and
\begin{equation}
\label{eq:Embed_Z_vs_R}
\left[1+\left(\frac{d\text{Z}}{d\text{R}}\right)^{2}\right]d\text{R}^2 
=
 \CF^4dr^2.
\end{equation}
Combining (\ref{eq:Embed_Z_vs_R}) and (\ref{eq:Embed_R_vs_r}) results in
\begin{equation}\label{eq:Embed_Z_vs_r}
\left(\frac{d\text{Z}}{dr}\right)^{2}=-4r\psi^2\frac{d\psi}{dr}\left(\psi+r\frac{d\psi}{dr}\right).
\end{equation}
Since the pseudo-spectral elliptic solver gives $\CF$ as a function of $r$, 
Eqs.~(\ref{eq:Embed_R_vs_r}) and~(\ref{eq:Embed_Z_vs_r}) allow us to solve for
the embedding radius R and the embedding height Z in terms of $r$.

\begin{figure}
\includegraphics[scale=0.52]{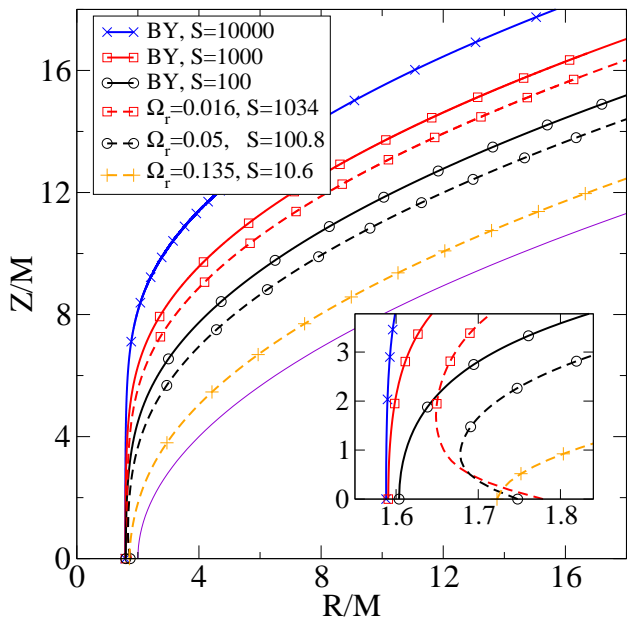}
\caption{\label{fig:PunctureXCTS_EmbedDiag}Embedding diagrams for puncture
  and quasiequilibrium initial data. Plotted is the embedding height Z
  as a function of the embedding radius R, both scaled by the mass $M$.
  For quasiequilibrium data (dashed lines), 
  Z=0 at $r=r_{\rm exc}$; for 
  puncture data (solid lines), 
  Z=0 at $r=R_{\rm inv}$.
  The thin solid purple curve represents the embedding of a plane 
  through a Schwarzschild black hole in Schwarzschild slicing.}
\end{figure}

Figure~\ref{fig:PunctureXCTS_EmbedDiag} shows embedding diagrams for
three sets of QE-XCTS and puncture data. We have set Z=0 at $r=r_{\rm
  exc}$ for QE-XCTS data and at $r=R_{\rm inv}$ for puncture data.
This figure also contains the embedding of a plane through Schwarzschild
in Schwarzschild coordinates (i.e. the $S=0$ limit of BY puncture data), given by
$R/M=Z^2/(8M^2)+2$.  
    Both puncture data and CFMS data exhibit a lengthening throat with
    increasing spin $S/M^2$.
For puncture data, this lengthening can be deduced from the
analytical results in Sec.~\ref{sec:SingleBHPunctureData}:
as the spin parameter $S$ of the
puncture data increases by a factor of 10 
while $m_p\equiv 1$ is held constant, we find from
Eq.~(\ref{eq:PunctureThroat-L}) that ${\cal L}/S^{1/2}$ should increase
by
\begin{equation}
\Delta{\cal L}/S^{1/2}=\frac{96^{1/4}}{2}\ln10 \approx 3.60,
\end{equation}
where the factor $1/2$ arises because $R_{\rm inv}=m_p/2=0.5$.
The embedding diagram shows only the top half of the throat,
and $S^{1/2}\approx M$ 
[cf. the discussion after Eq.~(\ref{eq:PunctureThroat-LC-Ratio})].  
Therefore in Fig.~\ref{fig:PunctureXCTS_EmbedDiag} 
the $S=100, 1000, 10000$ lines for BY (puncture) data 
should be spaced by $\Delta Z/M\approx 1.80$ for large 
$R/M$.  This indeed is the case. 

The CFMS
  datasets appear to scale proportionally to $\sqrt{S}$, 
  which is similar
  to the puncture data's behavior.  
  Furthermore, the CFMS initial data sets also develop
  a lengthening throat as $S$ becomes large (the effect is not as
  pronounced as for puncture data, owing to the smaller maximal $S$ we
  achieved.) Thus it appears that large spin CFMS data might be
  similar to large spin puncture data.
However,  the throats of the QE-XCTS data show a bulge near
the bottom, because for these data sets R actually decreases with $r$
in the immediate vicinity of $r_{\rm exc}$.  This is unlike the
puncture data, which very clearly exhibit cylindrical throats,
consistent with the discussion leading to
(\ref{eq:PunctureThroat-LC-Ratio}). 

%#######################################
\section{Binary-black-hole initial data with nearly-extremal spins} 
%#######################################
\label{sec:BBHdata}\label{sec:BBHData}
In this section, we construct binary-black-hole initial data with
rapid spins, confining our attention to the special case of spins
aligned with the orbital angular momentum.  In the limit of large
separation, binary-black-hole puncture initial data will behave like
two individual puncture initial data sets.  
Specifically, we expect that it should be possible to
  construct puncture binary-black-hole initial data with 
initial spins $\SMM(t=0)\lesssim
  0.98$, but the spins
will rapidly drop to $\SMM\lesssim 0.93$ as  the black holes settle down.
For this reason, and also because
puncture data is not well-suited to our
pseudospectral evolution code, we will
restrict our attention to binary black holes constructed with the
QE-XCTS approach.

As laid out in Table~\ref{tab:IDSummary},
we first construct a family (labelled CFMS) of standard
conformally-flat initial data on maximal slices;
then, we turn our attention to families (labelled SKS) of 
superposed Kerr-Schild initial data.  Finally, we construct 
a few individual SKS initial-data sets which we evolve in 
Sec.~\ref{sec:Evolutions}.  
All of the data sets 
represent equal-mass, equal-spin
black holes with spins parallel to the orbital angular momentum. 

In this section, unless otherwise indicated, 
all dimensionless spins are the 
 approximate-Killing-vector spin $\AKVSMM$ 
{(Appendix~\ref{sec:QuasilocalSpin}),  and
the subscript ``AKV'' will be suppressed for simplicity.

%%%%%%%%%%%%%%%%%%%%%%%%%%%%%%%%%%%%%%%%%%%%%%%%%%%%%%%%%%%%%%%%

\subsection{Conformally flat, maximal slicing data (CFMS)}
\label{sec:CFBBHID}
\begin{figure}
\includegraphics[width=3.4in]{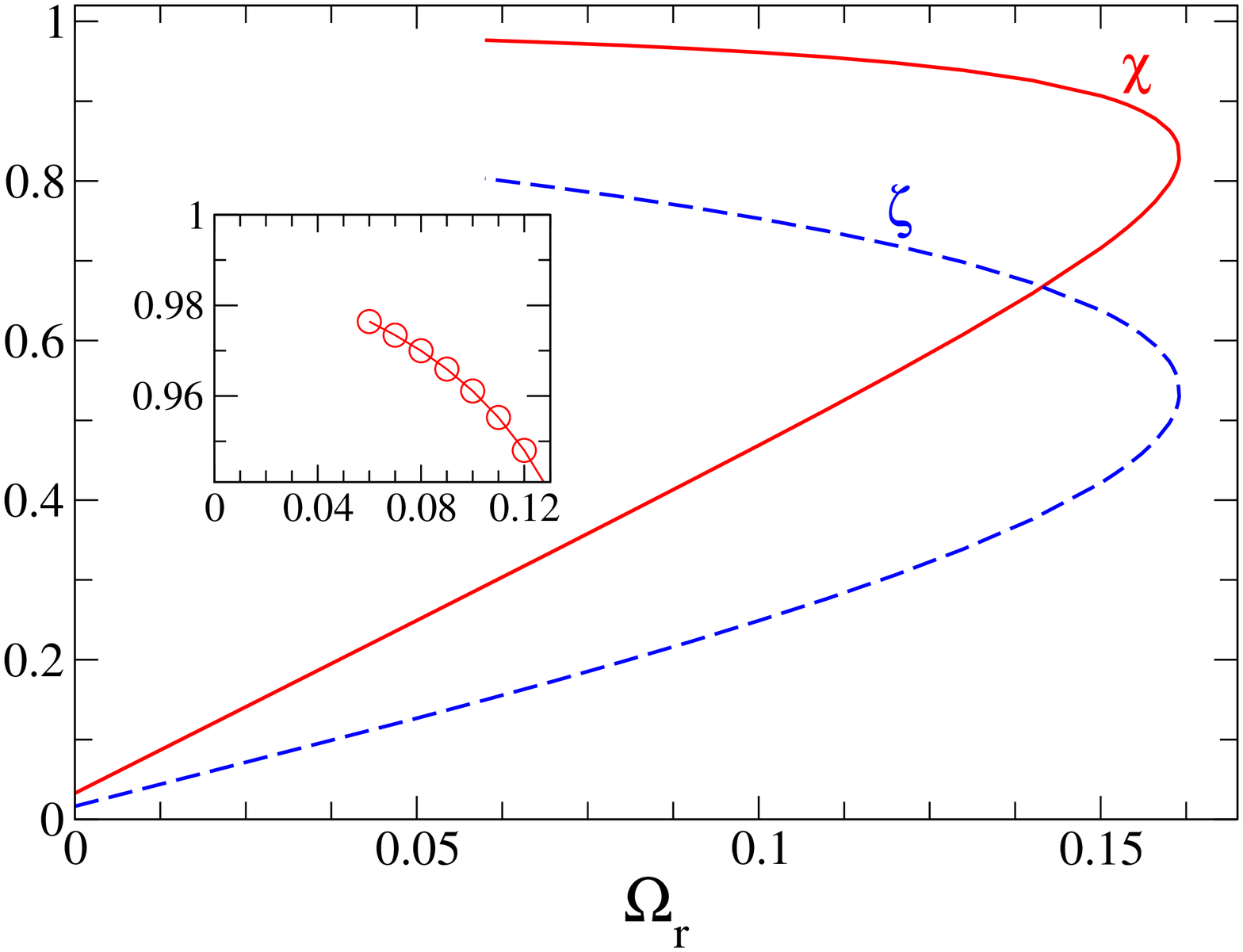}
\caption{
\emph{Main panel:}
Dimensionless spin $\SMM$ [Eq.~(\ref{eq:SMMDef})] and spin-extremality 
parameter $\SoTwoMirrMirr$ [Eq.~(\ref{eq:qDef})] 
for the family CFMS of spinning binary-black-hole initial data.
\emph{Inset:} Enlargement of $\SMM$ toward the end of the upper branch, 
with circles
denoting the individual initial data sets that were constructed.
Compare with Fig.~\ref{fig:CFBHMassSpin}.
\label{fig:NewMassSpin}\label{fig:BBH_CFMS_Dimensionless}}
\end{figure}
To construct conformally-flat binary-black-hole data, we solve the
same equations and boundary conditions as for the single-black-hole
case, as described in Sec.~\ref{sec:SingleBHQEXCTS}, 
with the main difference being
that we excise {\em two} spheres with radius $\rexc$ 
[cf. Eq.~(\ref{eq:rexc-XCTS-CF})] 
with centers on the x-axis at $x=\pm \CoordSep/2$.
The initial spins of the holes are set by adjusting $\Omega_r$, 
just as in the single-hole case.
The parameters $\mathbf{\Omega_0}$ and $\dot{a}_0$ in the outer
boundary condition on the shift [Eq.~\ref{eq:OutBCEnd}] determine the
initial angular and radial motion of the holes, which in turn
determine the initial eccentricity $e$ of the orbit. We set
$\mathbf{\Omega_0}=\Omega_0 \mathbf{e}_z$, where $\mathbf{e}_z$ is a
unit vector that points along the positive z axis. 
For the CFMS family of data sets considered here, we use 
values for $\Omega_0$ and $\dot{a}_0$ that should
result in closed, fairly circular orbits, since our choices of 
$\Omega_0$ and $\dot{a}_0$ lead to data sets that approximately satisfy 
the Komar-mass condition $\Eadm=M_K$ (cf.~\cite{Caudill-etal:2006}).
Specifically, on the lower branch of the resulting 
non-unique family of initial data,
\begin{eqnarray}\label{eq:KomarError}
\frac{\left|\Eadm -  M_K\right|}{\Eadm} \lesssim 1\%,
\end{eqnarray}
where the Komar mass is defined by (e.g., Eq.~(35) of 
Ref.~\cite{Caudill-etal:2006})
\begin{eqnarray}\label{eq:KomarDef}
M_K := \frac{1}{4\pi}\oint_\infty \left(\nabla_i \Lapse 
- \Shift^j \ExCurv_{ij} \right) \TwoAreaElement.
\end{eqnarray} (On the upper branch, $\Eadm$ and $M_K$ differ by 
up to 3\%.) 

As the rotation parameter $\Omega_r$ is varied 
(with the coordinate separation $d$ held fixed), 
we find that the CFMS-family 
of binary-black-hole initial data behaves qualitatively 
similarly to the analogous single-black-hole
initial data discussed
in Sec.~\ref{sec:SingleBHQEXCTS}. There is a maximal 
$\Omega_{r,\mathrm{crit}}$ such that no solutions can be found for 
$\Omega_r>\Omega_{r,\mathrm{crit}}$;  
for
values of $\Omega_r$ below $\Omega_{r,\mathrm{crit}}$, two solutions exist.
Figure~\ref{fig:NewMassSpin} plots the 
dimensionless  spin $\SMM$ and the spin-extremality parameter 
$\SoTwoMirrMirr$ 
against $\Omega_r$ for this family of initial data.
We only show values for one of the holes, 
since the masses and spins are equal. 
Spins larger than 
$\SMM\approx0.85$ appear on the upper branch. The highest spin we have been 
able to construct is larger than $\SMM=0.97$. 

%%%%%%%%%%%%%%%%%%%%%%%%%%%%%%%%%%%%%%%%%%%%%%%%%%%%%%%%%%%%%%%%

\subsection{Superposed-Kerr-Schild data}
\label{sec:SKSBBHID}

In this section, we solve the same equations and boundary conditions as in the
conformally flat case, except that we use SKS free data
(Sec.~\ref{sec:FormalismSKS}) instead of conformally-flat free data.
To construct the individual Kerr-Schild data, we need to choose 
for each black hole the coordinate location of its center, its conformal mass 
$\tilde{M}$,  conformal spin $\tilde{S}$, and its boost-velocity.  
We center the black holes on the x-axis at $x=\pm\CoordSep/2$,
use the same mass $\tilde{M}=1$ for both black holes, and
set the boost velocity to $(0,\pm \CoordSep\,\Omega_0/2, 0)$. 
The conformal spins are always equal and are aligned with the orbital 
angular momentum of the holes.

In contrast to the CFMS data, there are now 
\emph{two} parameters that influence
the black holes' spins: i) the rotation parameter $\Omega_r$ in 
Eq.~(\ref{eq:InnerShiftBC}), and ii) the conformal spin $\tilde{S}$.
For concreteness, we choose to construct
data for four different values of the 
conformal spin: $\tilde{S}/\tilde{M}^2=0, 0.5, 0.93,$ and $0.99$.
For each choice, we construct a family of initial data sets for 
different values of $\Omega_r$, which we label as
SKS-0.0, SKS-0.5, SKS-0.93, and SKS-0.99 respectively. 

Other choices that went into the
construction of the SKS initial data sets are as follows:
\begin{itemize}
\item 
The excision boundaries are chosen to be 
the coordinate
locations of the horizons of the individual Kerr-Schild metrics, i.e.
they are surfaces of constant Kerr-radius
\begin{eqnarray}
\rexc^{\rm Kerr} = \tilde{r}_+ := \tilde{M} + \sqrt{\tilde{M}^2-\tilde{S}^2},
\end{eqnarray} 
length-contracted by the Lorentz-factor appropriate for the boost velocity 
of each black hole.  This length-contraction accounts for 
the tangential motion of the hole but neglects the much smaller 
radial motion. 
\item When superposing the individual Kerr-Schild metrics, 
we use a damping length scale  $w=10\rexc^{\rm Kerr}$ 
[cf. Eqs.~(\ref{eq:CMetricSKS}) and~(\ref{eq:TrKSKS})],
except for the SKS-0.99 family, which uses $w=\CoordSep/3$.
\item The orbital frequency $\Omega_0$ and radial expansion
$\dot{a}_0$ are held fixed along each family. 
We expect that our choices for $\Omega_0$ and $\dot{a}_0$ will lead to 
bounded, fairly circular orbits, since
\begin{eqnarray}
\frac{\left|\Eadm-M_K\right|}{\Eadm} \lesssim 3\%.
\end{eqnarray}
In Sec.~\ref{sec:EccentricityReduction} we 
reduce the orbital eccentricity for one data set
in the family SKS-0.93.
\end{itemize}

%%%%%%%%%%%%%%%%%%%%%%%%%%%%%%%%%%%%%%%%%%%%%%%%%%%%%%%%%%%%%%%%
\begin{figure}
\includegraphics[width=3.2in]{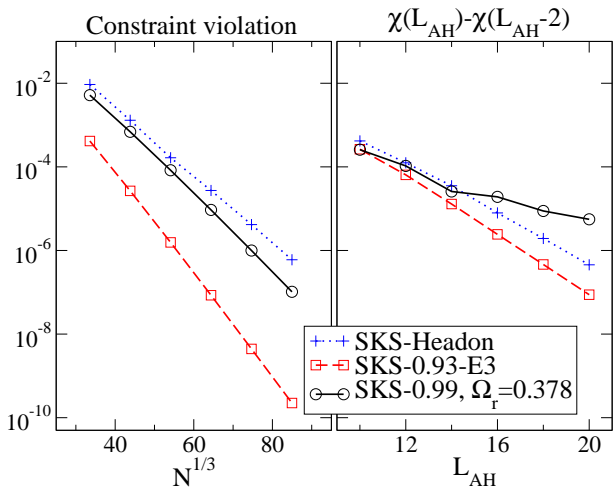}
\caption{
 Convergence of the 
spectral elliptic solver.  {\it Left panel}: The residual
constraint violation as a function of the total number of grid-points $N$
when running the elliptic solver at several different resolutions. 
{\it Right panel:} Convergence of the black hole dimensionless spin 
$\SMM$ [Eq.~(\ref{eq:SMMDef})]
with increasing resolution $L_{\rm AH}$ of the apparent horizon finder,
applied to the highest-resolution initial data set of the left panel. 
The three curves in each panel represent three different initial data sets:
One from the family SKS-0.99, as well as the two initial data sets that
are evolved in Sec.~\ref{sec:Evolutions}.
\label{fig:IDPlot_Convergence}
}
\end{figure}
%%%%%%%%%%%%%%%%%%%%%%%%%%%%%%%%%%%%%%%%%%%%%%%%%%%%%%%%%%%%%%%%

We again solve the XCTS equations using the spectral 
elliptic solver of Ref.~\cite{Pfeiffer2003};
the families of SKS initial data sets that we construct are summarized in 
Table~\ref{tab:IDSummary}.  The
elliptic solver needs some initial guess for the variables to be
solved for; we superpose the respective single-black hole
Kerr-Schild quantities, 
i.e.
\begin{subequations}
\begin{eqnarray}
\CF & = & 1,\\
\Lapse\CF & = & 1 + \sum_{a=1}^n e^{-r_a^2/w_a^2} (\alpha_a-1),\\ 
\Shift^i & = & \sum_{a=1}^n e^{-r_a^2/w_a^2} \beta^i_a,
\end{eqnarray}
\end{subequations} where $n=2$ and $\Lapse_a$ and $\beta^i_a$ are the lapse 
and shift corresponding 
to the boosted, spinning Kerr-Schild metrics
$g_{ij}^a$ used in the conformal metric $\CMetric_{ij}$.
Convergence of the elliptic solver and 
spin are demonstrated in Fig.~\ref{fig:IDPlot_Convergence} by 
showing the decreasing constraint 
violation\footnote{The constraint violation is
$\sqrt{\LTwoNorm{\Ham}^2+\LTwoNorm{\cal{\Mom}}^i \LTwoNorm{\cal{\Mom}}^j
\delta_{ij}}$, where $\Ham$ and $\Mom^i$ are the residuals of  
Eqs.~(\ref{eq:Ham})--(\ref{eq:Mom}) and the L2 norm is given by 
Eq.~(\ref{eq:LTwoNorm}).}
and differences in spin with increasing resolution.

%%%%%%%%%%%%%%%%%%%%%%%%%%%%%%%%%%%%%%%%%%%%%%%%%%%%%%%%%%%%%%%%
\begin{figure}
\includegraphics[width=3.4in]{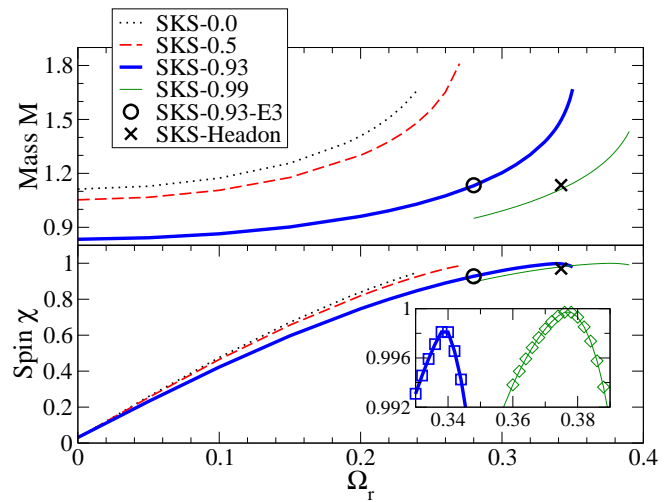}
\caption{
The mass $\HorizonMass$ (\emph{upper panel}) and 
dimensionless spin $\SMM$ (\emph{lower panel})
of one of the holes for Superposed-Kerr-Schild, 
binary-black-hole initial data sets with spins aligned with the 
orbital angular momentum. 
The mass and spin are plotted against $\Omega_r$ 
[Eq.~(\ref{eq:InnerShiftBC})] for four different 
choices of the conformal spin: $\tilde{S}=0$, $0.5$, $0.93$, and $0.99$. 
Also shown are the data sets~$\circEvLabel$---identical to the 
$\Omega_r=0.28\tilde{\HorizonMass},
\mbox{ }\tilde{S}=0.93\tilde{\HorizonMass^2}$ 
data set on the solid curve but with lower 
eccentricity---and $\plungeEvLabel$; both sets are 
evolved in  
Sec.~\ref{sec:Evolutions}.
The \emph{inset} in the lower panel shows a close-up of 
the spins 
as they approach unity, with symbols denoting the individual data sets.
\label{fig:IDPlot_SKS_MassSpin}}
\end{figure}
%%%%%%%%%%%%%%%%%%%%%%%%%%%%%%%%%%%%%%%%%%%%%%%%%%%%%%%%%%%%%%%%

We now turn our attention to the physical properties of the
SKS initial data sets. Figure~\ref{fig:IDPlot_SKS_MassSpin} shows the 
horizon mass $M$ and the dimensionless spin $\SMM$ 
of either black hole for the four 
families of SKS initial data. As expected, we find that 
generally the spin $\SMM$ increases with increasing $\Omega_r$.  
For each of the SKS-families,
we find that the elliptic solver fails to converge for sufficiently
large $\Omega_r$.  We suspect that the SKS-families exhibit a
turning point, similar to the CFMS-single and binary black hole
initial data shown in Figs.~\ref{fig:CFBHMassSpin}
and~\ref{fig:BBH_CFMS_Dimensionless}.  If this is the case, 
Fig.~\ref{fig:IDPlot_SKS_MassSpin} only shows the lower branch
of each family, and an additional branch of solutions will be present. 
Because we are satisfied with the spin magnitudes that are possible
along the lower branch, we 
do not attempt to find the upper branch here.  

In contrast to the CFMS data sets (where the lower branch only
allowed spins as large as $\SMM\lesssim 0.85$), the SKS initial data allows
spins that are quite close to unity.  
For the different SKS families, we are able 
to construct initial data with spins as large as
\begin{itemize}
\item $\SMM\approx 0.95$ for SKS-0,
\item $\SMM\approx 0.985$ for SKS-0.5,
\item $\SMM\approx 0.998$ for SKS-0.93,
\item $\SMM\approx 0.9997$ for SKS-0.99.
\end{itemize}
These spins are far closer to extremal than possible with Bowen-York
initial data [$\SMM\lesssim 0.984$ (Fig.~\ref{fig:puncture})] or
conformally flat, maximally sliced XCTS initial data [$\SMM\lesssim
0.85$ or $\lesssim 0.99$ along the lower and upper branch, respectively 
(Fig.~\ref{fig:CFBHMassSpin})].

We note that the spins in the SKS binary-black-hole
initial data families are only weakly
dependent on the orbital parameters $\Omega_0$ and $\dot a_0$.  
This can be seen from the individual data-point 
labeled SKS-0.93-E3 shown in
Fig.~\ref{fig:IDPlot_SKS_MassSpin}.  This  
data-set uses different values for $\Omega_0$ and $\dot a_0$ 
but is nevertheless close to the family SKS-0.93. 
The initial data sets SKS-0.93-E3 and SKS-HeadOn will
be discussed in detail in Sec.~\ref{sec:Evolutions}.

%%%%%%%%%%%%%%%%%%%%%%%%%%%%%%%%%%%%%%%%%%%%%%%%%%%%%%%%%%%%%%%%
\begin{figure}
\includegraphics[width=3.4in]{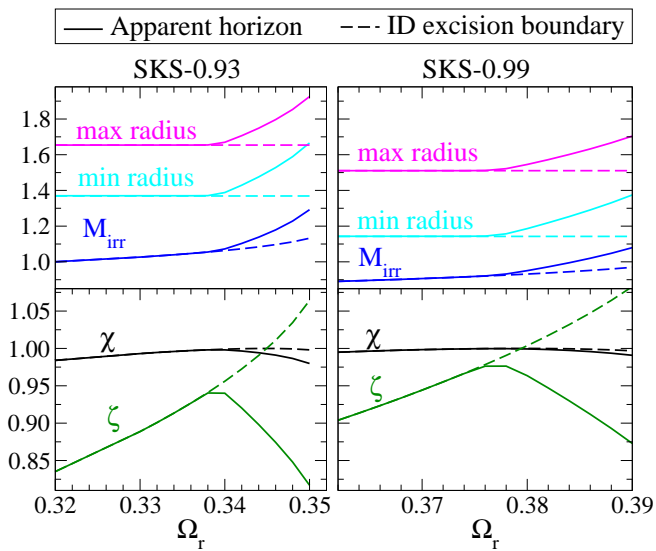}
\caption{
The irreducible mass $\Mirr$ and Euclidean coordinate radius $r$ 
(\emph{upper panels}) and dimensionless spin 
$\SMM:=\Spin/\HorizonMass^2$ and 
spin-extremality parameter $\SoTwoMirrMirr:=S/(2 \Mirr^2)$ 
(\emph{lower panels})
for one of the black holes in the 
SKS-0.93 (\emph{left}) and SKS-0.99 (\emph{right}) initial-data-set families. 
These quantities are computed on 
two surfaces: i) the apparent horizon (solid lines), and 
ii) the excision boundary of the 
initial data (dashed lines). 
Because we enforce that the excision surface is a 
marginally trapped surface, typically the apparent horizon and 
excision boundary coincide. However, if $\Omega_r$ is increased beyond the 
values where $\SMM$ approaches unity, the apparent horizon lies outside of 
the excision surface. The excision surface can obtain superextremal spins 
($\SoTwoMirrMirr>1$), but only when it is enclosed by a subextremal horizon. 
\label{fig:IDPlot_BBH_SKS_Maximal}}
\end{figure}
%%%%%%%%%%%%%%%%%%%%%%%%%%%%%%%%%%%%%%%%%%%%%%%%%%%%%%%%%%%%%%%%

The inset of Fig.~\ref{fig:IDPlot_SKS_MassSpin} highlights a
remarkable feature of the SKS-0.93 and SKS-0.99 families: with
increasing $\Omega_r$, the spin initially increases but eventually
{\em decreases}.  Figure~\ref{fig:IDPlot_BBH_SKS_Maximal} investigates
this behavior in more detail, where this
effect is more
clearly visible in the lower two panels: both the spin $\SMM$ and the
extremality parameter $\SoTwoMirrMirr$ of the apparent horizon 
change direction and begin to decrease. For $\Omega_r$ smaller than this
critical value, the apparent horizon finder always converges onto
the excision surfaces, which by virtue of the boundary condition
Eq.~(\ref{eq:AH-BC}), are guaranteed to be marginally trapped
surfaces.  As $\Omega_r$ is increased through the critical value (at
which $\SMM$ and $\SoTwoMirrMirr$ change direction), a {\em second}
marginally trapped surface (solid line) splits off from 
the excision surface (dashed line) and
moves continuously outward.  This can be seen in
the upper panels of Fig.~\ref{fig:IDPlot_BBH_SKS_Maximal}, which plot
the minimal and maximal coordinate radius and the irreducible mass
of both the excision surface and the 
outermost marginally trapped surface, which is by definition the apparent 
horizon. 

But what about the excision surface?  The boundary condition
Eq.~(\ref{eq:AH-BC}) forces the excision surface to be a marginally
trapped surface, independent of the value of $\Omega_r$. For 
sufficiently large $\Omega_r$, however, the excision surface is surrounded 
by a larger marginally trapped surface and thus is \emph{not}
the apparent horizon.  The dashed lines in
Fig.~\ref{fig:IDPlot_BBH_SKS_Maximal} present data for the excision
surface.  These lines continue smoothly across the point where the
second marginally trapped surface forms.  The extremality parameter
$\SoTwoMirrMirr$ for the excision surface continues to increase and
eventually becomes larger than unity;  the excision surface can then
be thought of as having a superextremal spin.  However, for the outer
marginally trapped surface---the true apparent horizon---the 
extremality parameter always satisfies $\SoTwoMirrMirr<1$.  The
irreducible mass $\Mirr$ of this surface increases faster than the
spin, and therefore $\SoTwoMirrMirr=S/(2\Mirr^2)$ decreases with
increasing $\Omega_r$.  

One might interpret these results as support
of the cosmic censorship conjecture. The XCTS boundary conditions
(\ref{eq:AH-BC}) and (\ref{eq:InnerShiftBC}) control the location
and the spin of the excision surface.  By appropriate choices for
the shift boundary condition (\ref{eq:InnerShiftBC}), we can force
the excision surface to become superextremal.  However, before this
can happen, a new horizon appears, surrounding the excision surface
and hiding it from ``our'' asymptotically
flat end of the spacetime.  The newly formed outer horizon always
remains subextremal. 

%%%%%%%%%%%%%%%%%%%%%%%%%%%%%%%%%%%%%%%%%%%%%%%%%%%%%%%%%%%%%%%%
\subsection{Suitability for evolutions}
%%%%%%%%%%%%%%%%%%%%%%%%%%%%%%%%%%%%%%%%%%%%%%%%%%%%%%%%%%%%%%%%

%%%%%%%%%%%%%%%%%%%%%%%%%%%%%%%%%%%%%%%%%%%%%%%%%%%%%%%%%%%%%%%%
\begin{figure}
\includegraphics[width=3.4in]{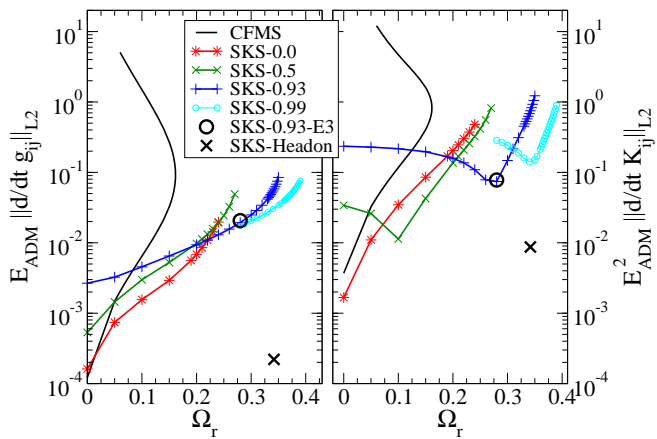}
\caption{\label{fig:TimeDerivatives}
  The time derivatives
  of the metric (\emph{left panel}) and extrinsic curvature 
  (\emph{right panel}).
  In the superposed-Kerr-Schild (SKS) data sets, 
  $\LTwoNorm{\partial_t K_{ij}}$
  has minima near values of $\Omega_r$ for which the 
  dimensionless spin $\SMM$ 
  is approximately equal to 
  the spin $\tilde{\Spin}$ of the conformal metric
  (cf. Fig.~\ref{fig:IDPlot_SKS_MassSpin}).  On the upper branch of the
  conformally-flat, maximally-sliced (CFMS) excision data, where the spin is
  $\SMM>0.83$ (Fig.~\ref{fig:NewMassSpin}), the time derivatives
  become much larger than the SKS time derivatives.
  The data sets~$\circEvLabel$ (with $\SMM\approx\tilde{S}=0.93$)
  and $\plungeEvLabel$ (with $\SMM\approx\tilde{S}=0.97$)
  are evolved in Secs.~\ref{sec:evCirc}--\ref{sec:evPlunge};
  the time derivatives are significantly lower for the set $\plungeEvLabel$ 
  because of the larger coordinate separation of the 
  holes ($d=100$ vs. $d=32$).
  \label{fig:TimeDerivs}}
\end{figure}
%%%%%%%%%%%%%%%%%%%%%%%%%%%%%%%%%%%%%%%%%%%%%%%%%%%%%%%%%%%%%%%%

In the previous sections, we have constructed a wide variety of 
binary-black-hole initial
data sets.  To get some indication about how suitable these are for
evolutions, we consider the initial time-derivatives of these 
data sets, $\partial_t g_{ij}$ and $\partial_t K_{ij}$. Recall
that solutions of the XCTS equations give a preferred initial lapse and
shift for the evolution of the initial data; hence, the time derivatives 
$\partial_t g_{ij}$ and $\partial_t K_{ij}$ can be computed by simply 
substituting the initial data into the ADM evolution equations.  
We expect initial data with smaller time-derivatives to be closer to 
quasi-equilibrium and to have less initial spurious radiation.

Figure~\ref{fig:TimeDerivs} presents the L2 norms of the time
derivatives, $\LTwoNorm{\partial_t
\SMetric_{ij}}$ and $\LTwoNorm{\partial_t \ExCurv_{ij}}$ where the L2
norm of a tensor $T_{ijk\cdots}(x)$ evaluated at $N$ gridpoints $x_i$
is defined as 
\begin{eqnarray}\label{eq:LTwoNorm}
\LTwoNorm{T_{ijk\cdots}}:=\sqrt{\frac{1}{N}\sum_{i=0}^N \bar{T}^2(x_i)},
\end{eqnarray} where
\begin{eqnarray} 
\bar{T}:=\sqrt{T_{ijk\cdots}T_{i^\prime j^\prime k^\prime \cdots} 
\delta^{ii\prime} \delta^{jj\prime} \delta^{kk\prime} \cdots}.
\end{eqnarray}

Figure~\ref{fig:TimeDerivs} shows that generally
$\partial_tK_{ij}$ is larger than $\partial_t g_{ij}$.  This has
also been found in previous work, e.g.~\cite{Pfeiffer2003a}, and is
not surprising, because the XCTS formalism allows some control over
the time derivative of the metric through the free data $\tilde
u_{ij}= \partial_t\tilde g_{ij}$, whereas there is less control of 
$\partial_tK_{ij}$.
We note that for CFMS data, the time derivatives are larger
and grow more rapidly with $\SMM$ than
for SKS data; in particular, the time derivatives on the upper branch
are $\sim 10$ times larger than for SKS-initial data, 
suggesting that these data are 
much farther from equilibrium.

In the SKS case, the time derivatives of $\ExCurv_{ij}$
have local minima at particular values of $\Omega_r$;
comparison with Fig.~\ref{fig:IDPlot_SKS_MassSpin} 
gives spins $\SMM$ at these minima of 
$\LTwoNorm{\partial_tK_{ij}}$ as follows:
\begin{itemize}
\item SKS-0.5: $\Omega_r\approx 0.1$, $\SMM\approx 0.45$,
\item SKS-0.93: $\Omega_r\approx 0.28$, $\SMM\approx 0.93$,
\item SKS-0.99: $\Omega_r\approx 0.34$, $\SMM\approx 0.98$.
\end{itemize}
Note that these minima occur at values of
$\Omega_r$ such that $\SMM \approx \tilde{\Spin}/\tilde{M}^2$;
that is, transients in the initial data and presumably the spurious
radiation are minimized when the conformal spin and AKV spin are
consistent. For this reason, we conclude that SKS initial data 
with $\SMM\approx \tilde S/\tilde M^2$ is preferable; this is the type
of initial data we will evolve in the next section.

Also note that minimizing the spurious radiation has
purely numerical advantages: the spurious radiation typically
has finer structure (and thus requires higher resolution) that the
physical radiation.  If such radiation is minimized, 
the numerical evolutions may require less resolution and will be more
efficient. Conformally-curved initial data has been found to reduce the 
amount of spurious radiation in Refs.~\cite{Hannam2007b,Thesis:Lovelace}.

%###############################################
\section{Exploratory evolutions of superposed Kerr-Schild (SKS) 
initial~data}
%###############################################
\label{sec:Evolutions}

So far, we have confined our discussion to 
black hole spins in the {\em initial data}. 
In this section, we compare the initial spin to the value 
to which the spin relaxes after the initial 
burst of spurious radiation, when the holes have settled down. 
Recall, for instance, that for Bowen-York puncture initial data 
with spins close to the maximal possible value [$\SMM(t=0)\approx 0.98$], 
the spins quickly relax by
about $\Delta\SMM\approx 0.05$ to a maximal possible relaxed value
of $\SMM(\trelax)\approx 0.93$ (cf.~\cite{DainEtAl:2008}).
While the SKS data presented in Sec.~\ref{sec:SKSBBHID} can 
achieve larger initial spins [$\SMM(t=0)=0.9997$] than 
conformally-flat puncture data, only evolutions can determine
$\Delta\SMM$ and $\SMM(\trelax)$.

Therefore, in this section we perform brief, exploratory evolutions of 
some SKS initial data sets to determine $\Delta\SMM$ for those 
data sets.\footnote{ 
Note that there is no universal value of $\Delta\SMM$---it will differ 
for different initial data sets, even within the same
family of initial data.}
Besides determination of $\SMM(\trelax)$, these evolutions  
will also allow us to demonstrate that 
the technique of eccentricity reduction developed in 
Ref.~\cite{Pfeiffer-Brown-etal:2007} is applicable to SKS initial data
as well as to compare the spin measures defined in 
Appendices~\ref{sec:spin} and~\ref{sec:SpinFromShape}. 
The focus here lies 
on initial data, and we evolve only
long enough for our purposes.
Longer simulations that 
continue through merger and ringdown are the subject of ongoing research.

This section is organized as follows. 
In Sec.~\ref{sec:EvDetails}, we 
summarize the evolution code that we will use. 
In Sec.~\ref{sec:EccentricityReduction}, we 
perform eccentricity reduction on 
one of the data sets in the SKS-0.93 family, which corresponds to 
an orbiting binary black hole with equal masses and equal spins 
(of magnitude $\SMM\approx0.93$) aligned with the orbital angular momentum.
Then, in Sec.~\ref{sec:evCirc}, we
evolve the resulting low-eccentricity data set (labeled SKS-0.93-E3).
Finally, In Sec.~\ref{sec:evPlunge}, we 
evolve a head-on plunge of SKS initial data (labeled SKS-Headon)
representing
two widely-separated black holes with initial spins 
of magnitude $\SMM=0.970$ and direction normal to the equatorial plane.

%%%%%%%%%%%%%%%%%%%%%%%%%%%%%%%%%%%%%%%%%%%%%%%%%%%%%%%%%%%%%%%%
\subsection{Description of evolution code}
\label{sec:EvDetails}
%%%%%%%%%%%%%%%%%%%%%%%%%%%%%%%%%%%%%%%%%%%%%%%%%%%%%%%%%%%%%%%%

The initial data are evolved using the Caltech-Cornell pseudospectral
evolution code {\tt SpEC}~\cite{Scheel2006}.  The details of the
evolution methods, equations, and boundary conditions that we use are
the same as those described in Ref.~\cite{Boyle2007}.  
The singularities are excised, with the excision surfaces chosen to lie 
slightly inside the black hole horizons.
Note that whereas Ref.~\cite{Boyle2007} excises
coordinate spheres inside the black holes' apparent horizons, here we use
Lorentz-contracted ellipsoidal excision boundaries 
which are adapted to the shape of the initial 
apparent horizons.

The highest-resolution initial data set 
(with $N\approx 85^3$ gridpoints) is interpolated onto evolution grids 
labelled $N1$, $N2$, and $N3$ with
approximately $61^3$, $67^3$, and $74^3$ gridpoints, respectively.
The outer boundary is at a coordinate radius of $r=32d$ for the
orbiting simulation discussed in Secs.~\ref{sec:EccentricityReduction}
and~\ref{sec:evCirc}  and at $r=14d$ for the head-on
simulations discussed in Sec.~\ref{sec:evPlunge}.  This translates to
about $r=450\Eadm$ and $r=620\Eadm$ for the
orbiting and head-on simulations, respectively.
As in earlier 
simulations~\cite{Scheel2006,Pfeiffer-Brown-etal:2007,Boyle2007}, 
a small region of the evolution grid lies
inside the horizon and is not covered by the initial data grid; we 
extrapolate $\CF$, $\Lapse\CF$, and 
$\Shift^i$ into this region and then compute $g_{ij}$ and $K_{ij}$.

%%%%%%%%%%%%%%%%%%%%%%%%%%%%%%%%%%%%%%%%%%%%%%%%%%%%%%%%%%%%%%%%
\subsection{Eccentricity removal for orbiting SKS-binaries}
\label{sec:EccentricityReduction}
%%%%%%%%%%%%%%%%%%%%%%%%%%%%%%%%%%%%%%%%%%%%%%%%%%%%%%%%%%%%%%%%
We obtain initial data with small orbital eccentricity using 
the iterative method of 
Ref.~\cite{Pfeiffer-Brown-etal:2007}, as refined in
Ref.~\cite{Boyle2007}, applied here for the first time 
to binary-black-hole data with rapid spin.
In this method, the choice of 
$\Omega_0$ and $\dot{a}_0$ for the next iteration are made so that if 
the orbit were Newtonian, the eccentricity would vanish. 
For the non-Newtonian orbit here, successive iterations succeed in reducing 
the orbital eccentricity. 

This procedure is based on
 the proper separation $s$ between the apparent horizons, measured along
a coordinate line connecting the geometric centers of the apparent horizons.
The time derivative 
$d\ProperSep/dt$ is 
fitted to a five-parameter curve that, together with the 
initial proper separation $s(t=0)$ is used to define the eccentricity $e$
and to define improved values for 
$\Omega_0$ and $\dot a_0$.
Specifically,
\begin{subequations}\label{eq:EccentricityReduction}
\begin{eqnarray}
\frac{ds}{dt} & := & A_0 + A_1 t + B \cos\left(\omega t + \varphi\right),\\
e & := & \frac{B}{\omega s(t=0)},\\
\Omega_{0,\rm new} & := & \Omega_0 + \frac{B\sin\phi}{2\; s(t=0)},\\
\dot{a}_{0, \rm new} & := & \dot{a}_0 - \frac{B\cos\phi}{s(t=0)}
\end{eqnarray}
\end{subequations} 
Heuristically, the eccentricity is embodied by the 
oscillating part of $ds/dt$. 

%%%%%%%%%%%%%%%%%%%%%%%%%%%%%%%%%%%%%%%%%%%%%%%%%%%%%%%%%%%%%%%%
\begin{figure}
\includegraphics[width=3.4in]{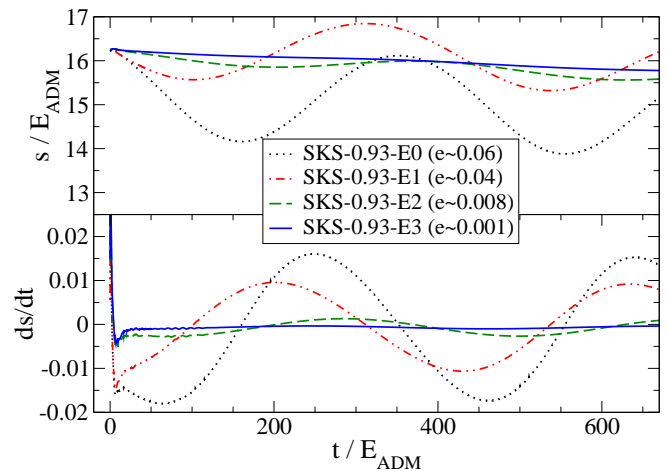}
\caption{\emph{Color online.} Eccentricity reduction for evolutions of 
superposed-Kerr-Schild binary-black-hole initial data. 
The proper separation $s$ (\emph{upper panel}) and 
its time derivative $ds/dt$ (\emph{lower panel}) are plotted
for initial data sets SKS-0.93-E0, -E1, -E2, and -E3, which
have successively smaller eccentricities $e$. 
All evolutions are performed at resolution $N1$. 
\label{evFig:dsdtVsTimeBothInMadmUnits}
}
\end{figure}
%%%%%%%%%%%%%%%%%%%%%%%%%%%%%%%%%%%%%%%%%%%%%%%%%%%%%%%%%%%%%%%%

Figure~\ref{evFig:dsdtVsTimeBothInMadmUnits} 
illustrates the eccentricity reduction for one of the 
data sets in family SKS-0.93. Plotted are the proper separation 
$s$ and its derivative $ds/dt$ for evolutions of
several initial data sets (summarized in Table~\ref{tab:IDSummary}): 
\begin{itemize}
\item set SKS-0.93-E0, which is identical to the set 
in family SKS-0.93 with $\Omega_r=0.28$ (Fig.~\ref{fig:IDPlot_SKS_MassSpin}); 
\item set SKS-0.93-E1, which is the same as SKS-0.93-E0 except that 
the orbital frequency $\Omega_0$ is manually adjusted to lower the 
orbital eccentricity somewhat; and
\item sets SKS-0.93-E2 and 
SKS-0.93-E3, which are successive iterations (starting from set SKS-0.93-E1)  
of the eccentricity-reduction scheme 
Eqs.~(\ref{eq:EccentricityReduction}).
\end{itemize}
The \emph{ad hoc} adjustment of $\Omega_0$ was somewhat effective,
  reducing $e$ by about 50\%.  The subsequent iterations using
  Eqs.~(\ref{eq:EccentricityReduction}) reduced $e$ by 
   factors of about 5
  and 8, respectively.  Surprisingly, the lowest eccentricity,
  corresponding to a smooth inspiral trajectory is obtained with a
  positive $\dot a_0=3.332\times 10^{-4}$. 
  This is not due to insufficient resolution; for 
  SKS-0.93-E3, we have verified that we obtain 
  the same eccentricity $e\sim0.001$ for all three numerical resolutions 
  N1, N2, N3.

Note that we choose to stop 
the evolutions at about $t=670\Eadm$, which 
corresponds to about 1.9 orbits; this is sufficient for reducing the 
eccentricity and for measuring $\Delta\SMM$.
In the next subsection, we discuss the evolution of the low-eccentricity set
SKS-0.93-E3 in detail, focusing on the relaxation of the spin $\SMM$.

%%%%%%%%%%%%%%%%%%%%%%%%%%%%%%%%%%%%%%%%%%%%%%%%%%%%%%%%%%%%%%%%
\subsection{Low-eccentricity inspiral with \boldmath$\SMM\approx0.93$ 
}
\label{sec:evCirc}

%%%%%%%%%%%%%%%%%%%%%%%%%%%%%%%%%%%%%%%%%%%%%%%%%%%%%%%%%%%%%%%%
\begin{figure}
\includegraphics[width=3.4in]{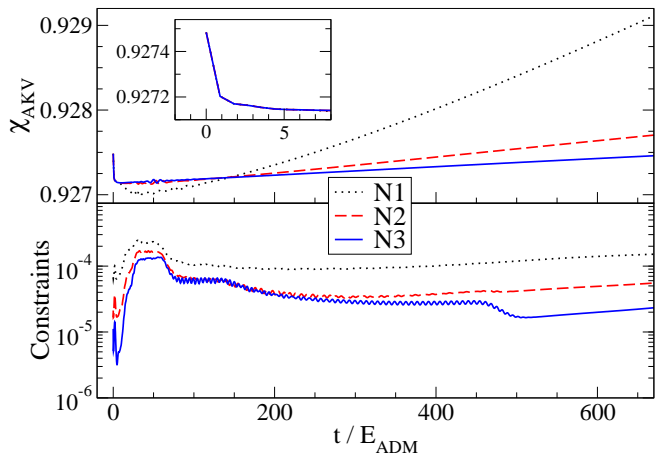}
\caption{
Convergence test of the evolution of 
the initial data set~$\circEvLabel$. Shown are evolutions on three different 
resolutions, $N1$, $N2$, and $N3$, with $N3$ being the highest resolution.
The \emph{top panel} shows
the approximate-Killing-vector (AKV) spin 
of one of the holes as a function of time, with the 
\emph{top inset} showing the spin's initial relaxation;
the \emph{bottom panel} shows the constraint violation 
as a function of time.
\label{evFig:SpinAndGhCe}}
\end{figure}
%%%%%%%%%%%%%%%%%%%%%%%%%%%%%%%%%%%%%%%%%%%%%%%%%%%%%%%%%%%%%%%%

We evolved the data-set SKS-0.93-E3 
at three different numerical resolutions for a duration of about
 $670\Eadm$, corresponding to about 1.9 orbits.
From post-Newtonian theory~\cite{kidder95}, we
estimate that this simulation would proceed through about 20 
orbits to merger.

Figure~\ref{evFig:SpinAndGhCe} presents a convergence test for
  this run.
The lower panel of Fig.~\ref{evFig:SpinAndGhCe} shows the
normalized constraint violation 
 (see Eq.~(71) of Ref.~\cite{Lindblom2006} for the precise definition.)
While the constraints are small, the convergence seems poor 
until $t\approx 500 \Eadm$.  For this time-period
the constraint violations at high resolution $N3$ are
dominated by the outgoing pulse of spurious 
radiation---i.e. far away from the black holes---which 
we have not attempted to adequately resolve. 
At $t\approx 500\Eadm$, the pulse of spurious radiation 
leaves the computational domain through the outer boundary;
afterwards, the constraints decrease 
exponentially with increasing resolution, as expected.

The upper panel of Figure~\ref{evFig:SpinAndGhCe} shows the AKV spin
$\AKVSMM=\Spin/\HorizonMass^2$ for the
  three runs with different resolutions $N1$, $N2$, and
$N3$. Based on the difference between $N2$ and $N3$, the spin
  of the evolution $N3$ should be accurate to a few parts in $10^{4}$.
  For the time-interval $5<t/\Eadm<670$,
 the measured spin on resolution $N3$ is
  consistent with begin constant  within its estimated accuracy. 
  Very early in the simulation,
  $t<5\Eadm$, the spin $\SMM$ changes 
  {\em convergently resolved} from its initial value $\chi(t=0)=0.927\,48$ 
  to a relaxed value $\SMM(\trelax)=0.927\,14$ 
  (see inset of Fig.~\ref{evFig:SpinAndGhCe}).
Therefore, for SKS-0.93-E3, we find $\Delta\SMM=0.000\,34$.

Contrast this result with the evolution of 
a binary black hole puncture initial data set with
large spins, which is reported in 
Ref.~\cite{DainEtAl:2008}: 
for that particular evolution,
 $\SMM(t=0)=0.967$, $\SMM(\trelax)=0.924$, i.e. $\Delta\SMM=0.043$, 
more than a factor 100 larger than for the evolution of SKS-0.93-E3 
reported here.
This comparison is somewhat biased against the puncture evolution 
in~\cite{DainEtAl:2008}, which starts at a smaller separation possibly 
resulting in 
larger initial transients.
However, even in the 
limit that the black holes are infinitely separated (i.e., in the 
single-black-hole limit), the spins in Bowen-York puncture data relax to 
values near $\varepsilon_J=J_{ADM}/E_{ADM}^2$; 
to achieve a final spin of
$\chi(\trelax)\approx 0.93$, the initial spin of Bowen-York data must be 
$\SMM(t=0)\approx 0.98$ (cf. Fig.~2 of Ref.~\cite{DainEtAl:2008}). 
We conclude that the spin relaxes by a much smaller amount in the SKS case 
than in Bowen-York puncture or inversion symmetric data.

%%%%%%%%%%%%%%%%%%%%%%%%%%%%%%%%%%%%%%%%%%%%%%%%%%%%%%%%%%%%%%%%
\begin{figure}
\includegraphics[width=3.4in]{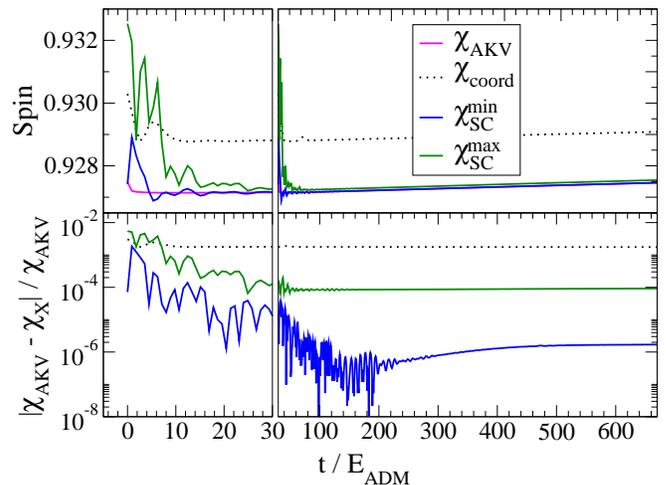}
\caption{ A comparison of different definitions of the 
spin. The \emph{top panel} shows the spin as a function of time for 
several different measures of the spin; the \emph{bottom panel}
shows the fractional difference between 
$\SMM_{\rm AKV}$ and alternative spin definitions.  
Note that for $t<30\Eadm$, the time-axis
has a different scaling to make the initial transients visible.
 \label{evFig:SpinDiffAndDiffSpinPaper}}
\end{figure}
%%%%%%%%%%%%%%%%%%%%%%%%%%%%%%%%%%%%%%%%%%%%%%%%%%%%%%%%%%%%%%%%

Figure~\ref{evFig:SpinAndGhCe} and the discussion in the
  previous paragraph only addresses the behavior of the AKV spin,
  where the approximate Killing vectors are computed from the
  minimization problem [cf. Eq.~(\ref{e:action})].  We now compare the
  different spin-definitions we present in
  Appendices~\ref{sec:QuasilocalSpin} and~\ref{sec:SpinFromShape}.
 Figure~\ref{evFig:SpinDiffAndDiffSpinPaper} compares these 
different definitions of 
the black hole spin for the $N3$ evolution of initial data set SKS-0.93-E3.
Shown are the AKV spin 
of one hole in the binary,
the scalar curvature (SC) spins $\SpinFromShapeMin$ and
$\SpinFromShapeMax$ of Appendix~\ref{sec:SpinFromShape} 
[Eqs.~(\ref{eq:SpinMin}) and~(\ref{eq:SpinMax})],
and also the spin obtained by 
using Eq.~\ref{eq:Spin} with a coordinate rotation vector instead of 
an approximate Killing vector (which we call the ``coordinate spin'' here).
After the holes have relaxed, the SC spins track the 
AKV spin more closely than does the coordinate spin. 
However, during very early times, as the holes are relaxing 
and the 
horizon shape is very distorted, the SC spins show much larger 
variations.
Consequently, the SC spin 
is a poorer measure of the spin at early times than even the coordinate 
spin.

%%%%%%%%%%%%%%%%%%%%%%%%%%%%%%%%%%%%%%%%%%%%%%%%%%%%%%%%%%%%%%%%
\subsection{Head-on plunge with \boldmath$\SMM\approx 0.97$}
\label{sec:evPlunge}
%%%%%%%%%%%%%%%%%%%%%%%%%%%%%%%%%%%%%%%%%%%%%%%%%%%%%%%%%%%%%%%%

In the previous subsection, we have seen that for
SKS binary-black hole-initial data
with $\SMM=0.93$, the initial spins change by
only a few parts in $10^4$.  
A spin $\SMM\approx 0.93$ is roughly the largest possible 
equilibrium spin that 
is obtainable using standard conformally-flat, Bowen-York puncture data 
(cf. the discussion at the beginning of  Sec.~\ref{sec:Evolutions}). 
We now begin to explore 
binary-black-hole simulations with
spin-magnitudes that are not obtainable with Bowen-York initial data methods.

We construct and evolve SKS binary-black-hole data for a head-on
plunge of two equal mass black holes with spins of equal magnitude
$\SMM=0.97$ and with the spins orthogonal to the line connecting the
black holes.  This data set, labelled~$\plungeEvLabel$, is summarized
in Table~\ref{tab:IDSummary} and was briefly
  discussed in Sec.~\ref{sec:SKSBBHID}, cf.
  Figs.~\ref{fig:IDPlot_Convergence}, \ref{fig:IDPlot_SKS_MassSpin}
  and~\ref{fig:TimeDerivatives}.  As
for the orbiting evolution SKS-0.93-E3, we adjust the
rotation parameter $\Omega_r$ so that conformal spin
$\tilde{\Spin}/\tilde{\HorizonMass}^2$ and AKV spin $\SMM$ are
approximately equal.
Starting such a simulation at close separation
results in rapid coordinate motion of the apparent horizons during the 
first few $\Eadm$ of the evolution.  These motions are
currently difficult to track with our excision code; therefore,
we begin at a larger separation $d$ than we used in the 
nearly-circular data sets described previously. 

%%%%%%%%%%%%%%%%%%%%%%%%%%%%%%%%%%%%%%%%%%%%%%%%%%%%%%%%%%%%%%%%
\begin{figure}
\includegraphics[width=3.4in]{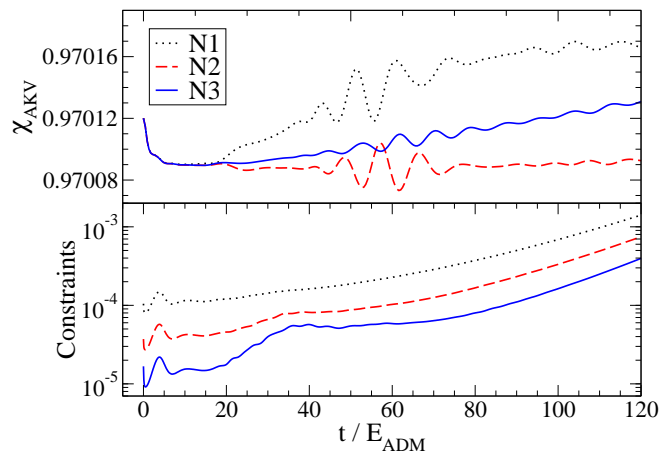}
\caption{
Convergence test of the head-on evolution
SKS-HeadOn.  Shown are evolutions at three different resolutions, 
$N1$, $N2$, and $N3$, with $N3$ being the highest-resolution.  
The {\em top panel} shows 
the approximate-Killing-vector (AKV) spin 
of one of the holes as a function of time;
the {\em bottom panel} shows the constraint violations as a function of time.
\label{fig:d100Convergence}}
\end{figure}
%%%%%%%%%%%%%%%%%%%%%%%%%%%%%%%%%%%%%%%%%%%%%%%%%%%%%%%%%%%%%%%%

Figure~\ref{fig:d100Convergence} 
presents a convergence test of the
constraints (lower panel) and the AKV spin $\SMM_{\rm AKV}$ (upper panel) 
during the subsequent evolution. Again, we are interested in the initial 
relaxation of the spins; therefore, we choose to stop evolution at 
$t\approx 120\Eadm$.  During this
time, the black hole proper separation decreased from $s(t=0)=47.6\Eadm$ 
to $s(t=120)=44.1\Eadm$.

During the first $\sim
10\Eadm$, $\SMM_{\rm AKV}$ shows (a numerically resolved) decrease of
about $3\times 10^{-5}$; this change arises due to initial
transients as the black holes and the full geometry of the
spacetime relax into an equilibrium configuration.
Subsequently, the spin remains constant to within about $10^{-4}$,
where these variations are dominated by numerical truncation error.

%%%%%%%%%%%%%%%%%%%%%%%%%%%%%%%%%%%%%%%%%%%%%%%%%%%%%%%%%%%%%%%%
\begin{figure}
\includegraphics[width=3.4in]{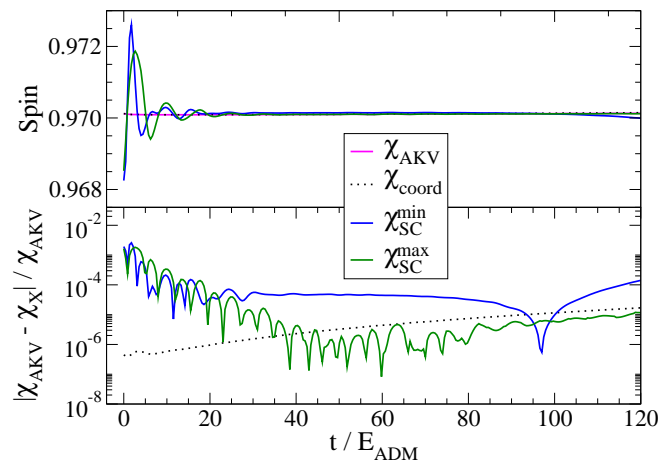}
\caption{
A comparison of various measures of the spin for
the head-on evolution of data set SKS-Headon, which is a
plunge of two equal-mass black holes with  
with parallel spins of magnitude $\AKVSMM=\Spin/\HorizonMass^2=0.970$ pointed 
normal to the equatorial plane. The \emph{top panel} shows various 
measures of the spin as a function of time, and the bottom panel 
shows the fractional difference between the 
approximate-Killing-vector (AKV) spin $\AKVSMM$ and alternative 
spin definitions.
\label{fig:d100SpinFromShape}}
\end{figure}
%%%%%%%%%%%%%%%%%%%%%%%%%%%%%%%%%%%%%%%%%%%%%%%%%%%%%%%%%%%%%%%%

Figure~\ref{fig:d100SpinFromShape} compares our various
spin-measures for the head-on simulation.  
Interestingly, the spin $\SMM_{\rm coord}$ computed from 
coordinate rotation vectors agrees 
much better with $\SMM_{\rm AKV}$ than for the SKS-0.93-E3 evolution, perhaps
because the black holes here are initially at rest.
The scalar-curvature 
(SC) spins $\SpinFromShapeMin$ and $\SpinFromShapeMax$, 
derived from the scalar curvature of the apparent horizon 
[Eqs.~(\ref{eq:SpinMin}) and~(\ref{eq:SpinMax})], show some
oscillations at early times; after the initial
relaxation, the SC spin agrees with the AKV spin 
to about 1 part in $10^4$. 

%#######################
\section{Discussion}
\label{sec:conclusions}
\subsection{Maximal possible spin}
%#######################

In this paper, we have examined a variety of methods for constructing
black hole initial data with a particular emphasis on the ability to
construct black holes with nearly-extremal spins. These are spins for which 
the dimensionless spin $\SMM=\Spin/\HorizonMass^2$ and 
spin-extremality parameter $\SoTwoMirrMirr=\Spin/(2\Mirr^2)$ are close to 
unity.

When discussing black hole spin, one needs to distinguish
between the \emph{initial} black hole spin and the \emph{relaxed} spin 
of the holes after they have settled down. Using conformally-flat 
Bowen-York (BY) data (both puncture data or inversion 
symmetric data) for single black holes, the largest obtainable spins are 
$\SMM\approx0.984,\SoTwoMirrMirr\approx 0.833$ 
(cf. Ref.~\cite{cook90} and Fig.~\ref{fig:puncture}).
With conformally-flat, maximally-sliced (CFMS), quasi-equilibrium 
extended-conformal-thin-sandwich (QE-XCTS)
data, 
we are able to obtain 
initial spins 
as large as 
$\SMM\approx 0.99, \SoTwoMirrMirr\approx 0.87$
for single black holes (Fig.~\ref{fig:CFBHMassSpin}). 
The limitations of BY puncture data and CFMS QE-XCTS data are already
present when constructing highly spinning single black holes;
therefore, we expect the methods to be able to construct binary-black-hole
data with similar spins as for single holes---i.e., up to 
about $0.98$. 
Construction of CFMS QE-XCTS binary-black-hole initial data
confirms this conjecture (compare
Fig.~\ref{fig:BBH_CFMS_Dimensionless} with Fig.~\ref{fig:CFBHMassSpin}).

For superposed-Kerr-Schild (SKS) initial data, 
the situation is different. For single black
holes,  SKS data reduce to the analytical Kerr solution, 
without any
limitations on the spin magnitude.  Thus limitations of SKS data will
only be visible for binary-black hole configurations.  As
Sections~\ref{sec:BBHdata} and~\ref{sec:Evolutions} show, however,
those limitations are quite minor. 
SKS data can indeed achieve initial spins 
that are much closer to extremality than what is possible with BY data 
or CFMS QE-XCTS data; 
we have explicitly demonstrated this by
constructing 
SKS data for binary black holes with
$\SMM\approx 0.9997,\SoTwoMirrMirr\approx 0.98$,
as can be seen from Figs.~\ref{fig:IDPlot_SKS_MassSpin} 
and~\ref{fig:IDPlot_BBH_SKS_Maximal}.

As the black hole spacetimes settle into equilibrium and emit spurious 
gravitational radiation, the initial spin $\SMM$ decreases to a smaller 
relaxed spin $\SMM(\trelax)$.
Thus an interesting quality factor for high-spin black hole initial data is
$\Delta\SMM=\SMM(t=0)-\SMM(\trelax)$ 
[Eq.~(\ref{eq:DiffSMM})]
considered as a function of the {\em relaxed} spin. 
The magnitude of $\Delta\SMM$ is
indicative of the amplitude of any initial transients, whereas the
maximally achievable $\SMM(\trelax)$ gives the largest possible
spin which can be evolved with such initial data. 
Figure~\ref{fig:DeltaChi} presents this plot, with the circle and
cross representing the two evolutions of SKS data which were 
described in 
Sec.~\ref{sec:Evolutions}.

%%%%%%%%%%%%%%%%%%%%%%%%%%%%%%%%%%%%%%%%%%%%%%%%%%%%%%%%%%%%%%%%
\begin{figure}
\includegraphics[width=3.2in]{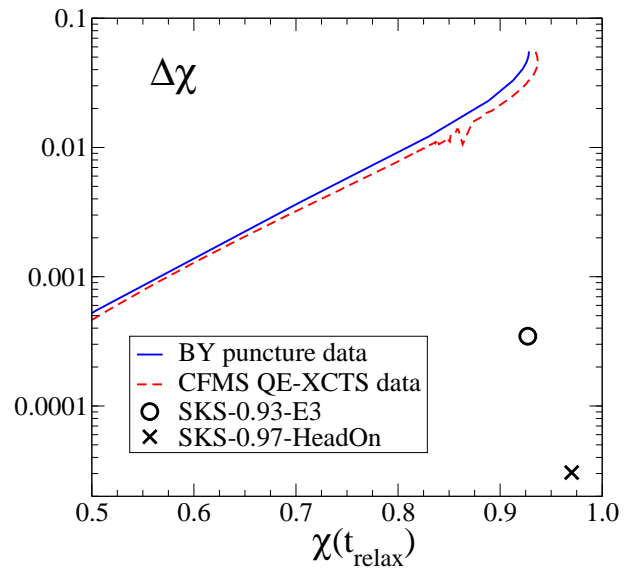}
\caption{
\label{fig:DeltaChi}
The change $\DiffSMM$ in black hole spin $\SMM$ during the initial relaxation 
of black hole initial data plotted as a function of the black-hole spin
{\em after} relaxation.  The SKS initial data constructed in this paper have
smaller transients and allow for larger relaxed spins.
}
\end{figure}
%%%%%%%%%%%%%%%%%%%%%%%%%%%%%%%%%%%%%%%%%%%%%%%%%%%%%%%%%%%%%%%%

We have not evolved high-spin puncture data, nor high-spin CFMS-XCTS data; 
therefore, we do not know precisely $\Delta\SMM$ for these initial data.
We estimate $\Delta\SMM$ for puncture data
by noting that evolutions of single-hole, BY puncture
data with large spins show~\cite{DainEtAl:2008} that
the black hole spin $\SMM:=S/M^2$ relaxes approximately to the
initial value of $\JEE:=\Jadm/\Eadm^2$. 
Therefore, 
for BY puncture data, we approximate
\begin{gather}\label{eq:BYDeltaChiStart}
\Delta\SMM\approx\varepsilon_J - \SMM(t=0),\\
\SMM(\trelax)\approx\varepsilon_J.\label{eq:BYDeltaChiEnd}
\end{gather}
This curve is plotted in Fig.~\ref{fig:DeltaChi}.  Because high-spin
single-black-hole,  CFMS QE-XCTS initial data and BY puncture data have
quite similar values of $\SMM(t=0)$ and $\varepsilon_J$, as well as 
similar embedding diagrams (cf. Fig.~\ref{fig:PunctureXCTS_EmbedDiag}),
we conjecture that Eqs.~(\ref{eq:BYDeltaChiStart})--(\ref{eq:BYDeltaChiEnd}) 
are also applicable to CFMS QE-XCTS
data. This estimate is also included in Fig.~\ref{fig:DeltaChi}. 
We see that both types of initial data result in a $\Delta\SMM$ of similar
magnitude which grows rapidly with $\SMM_{\rm relaxed}$. 

Perhaps the most remarkable result of Fig.~\ref{fig:DeltaChi}
is the extremely small change
in black hole spin during the relaxation of SKS initial data, even at
spins as large as $\SMM=0.97$. 
The small values of $\Delta\SMM$ combined with the ability to
construct initial data with initial spins $\SMM(t=0)$ as large as $0.9997$
(cf. Fig.~\ref{fig:IDPlot_SKS_MassSpin}) makes it highly likely that
SKS initial data are capable of constructing binary black holes
with relaxed spins significantly closer to unity than 0.97. 
Evolutions of initial data with spins $\SMM$ much closer to unity, 
i.e., farther into the regime that is inaccessible 
to conformally-flat data,
are a subject of our ongoing research.

In summary, the two main results of this paper are as follows:
\begin{itemize} 
\item SKS initial data can make binary black holes that initially have 
nearly-extremal spins, and
\item for SKS initial data, the relaxed spin is quite close to the 
initial spin, even when the spin is large.
\end{itemize}

\subsection{Additional results}

While working toward the main results discussed in the previous subsection, 
we have also established
several additional interesting results.  
We have considered spinning,
single-black-hole, puncture data which is identical to 
single-black-hole, spinning, inversion-symmetric data.  
Using this correspondence and
our accurate spectral elliptic solver, we revisited the relation
between black-hole spin $\SMM$, specific total angular momentum of the
space-time $\varepsilon_J$, and the spin-parameter $S$ for BY puncture data,
and established in Fig.~\ref{fig:puncture} that both $\SMM$ and
$\varepsilon_J$ approach their limits for $S\to\infty$ as 
{\em power-laws}, cf. Eqs.~(\ref{eq:puncture-powerlaw1}) 
and~(\ref{eq:puncture-powerlaw2})}.  We have also extended the analytical analysis of
Dain, Lousto, and Zlochower~\cite{DainEtAl:2008} of the throat region of 
high-spin
puncture data toward more quantitative results, including the precise
amplitudes of the conformal factor, throat circumference and throat
length, as well as their scaling with spin-parameter $S$ 
and puncture mass $m_p$
[Eqs.~(\ref{eq:Sqrt-r})--(\ref{eq:PunctureThroat-LC-Ratio})].
Furthermore, Ref.~\cite{DainEtAl:2008} implicitly assumed that the 
throat-region is
approximately spherically symmetric; our
Fig.~\ref{fig:PsiAngularStructure} presents explicit evidence
in support of this assumption, but also shows that the throat is not
precisely spherically symmetric.

We have also examined high-spin QE-XCTS initial data employing the 
common approximations of conformal
flatness and maximal slicing (CFMS).  With increasing angular frequency
$\Omega_r$ of the horizon, we discover non-unique solutions.  Thus,
the non-uniqueness of the XCTS equations can not only be triggered by
volume terms (as in~\cite{Pfeiffer-York:2005}) but also through boundary
conditions [in this case, by Eq.~(\ref{eq:InnerShiftBC})].  
Interestingly, CFMS QE-XCTS data appears to be very similar to BY
puncture data, in regard to nearly-extremal spins.  Both data formalisms
result in similar maximal values of $\SMM(t=0)$ and $\varepsilon_J$
(Figs.~\ref{fig:puncture}, \ref{fig:CFBHMassSpin}
and~\ref{fig:DeltaChi}) and have embedding diagrams which develop a
lengthening throat as the spin is increased 
(Fig.~\ref{fig:PunctureXCTS_EmbedDiag}).

We also have found an interesting property of the horizon geometries for 
SKS data, which one might interpret as support of the cosmic censorship 
conjecture. 
Specifically, we find that by increasing $\Omega_r$ sufficiently, 
we can in fact force the excision boundaries of the initial data to be 
``horizons'' (i.e. marginally trapped surfaces) with superextremal spin  
($\SoTwoMirrMirr > 1$). However, these superextremal surfaces are always 
enclosed by a larger, subextremal ($\SoTwoMirrMirr < 1$) apparent horizon. 

To measure black hole spins, we have employed and compared
several different techniques to measure black hole spin. Primarily, we use
a quasilocal spin definition based on (approximate) Killing vectors
[Eq.~(\ref{eq:Spin})].  This formula requires the choice of an
``approximate'' Killing vector, and we have used both straightforward
coordinate rotations to obtain $\SMM_{\rm coord}$ and solved Killing's
equation in a least-squares sense to obtain $\AKVSMM$
(see Appendix~\ref{sec:QuasilocalSpin} or details).
Furthermore, we introduced a new technique to define black-hole spin which
does not require choice of an approximate Killing vector and is
invariant under spatial coordinate transformations and 
transformations associated with the boost gauge ambiguity of the dynamical 
horizon formalism.  This new
technique is based on the extrema of the scalar curvature of the
apparent horizon.  Figures~\ref{evFig:SpinDiffAndDiffSpinPaper}
and~\ref{fig:d100SpinFromShape} show that all four spin measures agree
to good precision, but differences are noticeable. The spin-measures
based on the horizon curvature exhibit more pronounced variations
during the initial transients, and the quasilocal spin based on
coordinate rotations is off by several tenths of a percent.  The
quasilocal spin based on approximate Killing vectors $\SMM_{\rm AKV}$
has the smallest initial variations.

Finally, we would like to point out that a modified version of the
SKS-initial data has been very successfully used to construct 
black hole-neutron star initial data~\cite{FoucartEtAl:2008}.

\acknowledgments{
It is a pleasure to acknowledge useful discussions with 
Ivan Booth, Gregory Cook, Stephen Fairhurst, Lawrence Kidder, Lee Lindblom, 
Mark Scheel, 
Saul Teukolsky, and Kip Thorne. 
The numerical calculations in this paper were 
performed using the 
Spectral Einstein Code ({\tt SpEC}), 
which was primarily developed by Lawrence Kidder, 
Harald Pfeiffer, and Mark Scheel.  
We would also like to acknowledge the anonymous referee for reminding
us of an important technical caveat.
Some equations in this paper were obtained
using {\it Mathematica}.
This work was
supported in part by grants from the Sherman Fairchild Foundation to
Caltech and Cornell and from the Brinson Foundation to Caltech; by
NSF grants PHY-0652952, DMS-0553677,
PHY-0652929, and NASA grant NNG05GG51G at Cornell; and by
NSF grants PHY-0601459, PHY-0652995, DMS-0553302 and NASA grant
NNG05GG52G at Caltech.

}
\appendix

%%%%%%%%%%%%%%%%%%%%%%%%%%%%%%%%%%%%%%%%%%%%%%%%%%
%%%%%%%%%%%%%%%%%%%%%%%%%%%%%%%%%%%%%%%%%%%%%%%%%%%%%%%%%%%%%%%%
%\section{Defining the black hole spin}
\section{Quasilocal spin using approximate~Killing~vectors (AKV spin)}
\label{sec:QuasilocalSpin}
\label{sec:spin}

In this appendix and the one that follows, we address the task of 
defining the spin of a dynamical black hole, 
given $\SMetric_{ij}$ and $\ExCurv_{ij}$.
We use two different measures. The first, defined here, 
is a standard quasilocal angular momentum defined with 
approximate Killing vectors which correspond to approximate symmetries of 
a black hole's horizon. The second measure, defined in 
Appendix~\ref{sec:SpinFromShape}, infers the spin from 
geometrical properties 
(specifically, from the intrinsic scalar curvature $\HorizonShape$) of 
the apparent horizon, assuming that the horizon is that of a 
single black hole in equilibrium, (i.e., that the horizon is that of a 
Kerr black hole). 
Note that quantities relating to the geometry of 
the two-dimensional apparent horizon 
surface $\AH$ are denoted with a ring above them, to avoid confusion with the 
analogous quantities on the spatial slice, $\Slice$.

It has become standard in the numerical relativity community to compute the 
spin angular momentum of a black hole with the  
formula~\cite{BrownYork1993, Ashtekar2001, Ashtekar2003}
\begin{equation}
\Spin = \frac{1}{8 \pi}\oint_\AH \phi^i s^j K_{ij} \TwoAreaElement, 
\label{eq:Spin}
\end{equation}
where $s^i$ is the outgoing 
normal of $\AH$ embedded in $\Slice$ and $\vec \phi$ is an 
``azimuthal'' vector field, tangent to $\AH$.  The azimuthal vector 
field $\vec \phi$ carries information about the ``axis'' 
about which the spin is being computed.  There are, however, far more vector 
fields on a two sphere than there are axes in conventional Euclidean 
space.  We must find suitable criteria for fixing these azimuthal vector 
fields in numerical simulations, so that they reduce to the standard rotation 
generators when considered on a metric sphere.

Because angular momentum is generally thought of as a conserved charge 
associated with rotation symmetry---and indeed the quantity given in 
\eqref{eq:Spin} 
can be shown to be conserved under time evolution~\cite{BrownYork1993, 
Ashtekar2003} when $\vec \phi$ is a Killing vector of the dynamical horizon 
worldtube---it makes sense to 
consider Killing's equation to be the essential feature of the azimuthal 
vector field.  If a Killing vector on a dynamical horizon is tangent to each 
(two-dimensional) apparent horizon, then the vector field must be a Killing 
vector of each apparent horizon.  
However in a general spacetime, on an arbitrary apparent 
horizon, there is no reason to expect any Killing vectors to exist.  
So in the cases 
of most interest to numerical relativity, when there are no true rotation 
symmetries, we must relax the symmetry condition and find those vector fields 
that come ``closest'' to generating a symmetry of the apparent horizon.  
In other words, we seek optimal ``approximate Killing vectors'' of the 
apparent horizon.  

In~\cite{Dreyer2003}, a practical method for computing approximate 
Killing vectors was introduced, which has since been applied on numerous 
occasions, e.g.~\cite{Schnetter2006, Caudill-etal:2006, Campanelli2006d, 
Campanelli2007b} 
This method involves integrating 
the Killing transport equations along a predetermined network of coordinate 
paths.  The resulting vector field is guaranteed to be a Killing vector 
field if such a field exists and coincides with the computed field at any 
point on the network.  However if no true Killing field exists, the integral 
of the Killing transport equations becomes path dependent. 
This means that 
the computed vector field will depend in an essential way on the network of 
paths chosen for the integral.  Perhaps even more serious, 
if there is no true 
Killing field, then the transport of a vector around a closed path will not 
necessarily be an identity map.  As a result, the computed vector field 
cannot be expected to reduce to any smooth vector field in the limit that the 
network becomes more refined.  This kind of approximate Killing vector field 
is simply not mathematically well-defined in the continuum limit.

Here we will describe a kind of approximate Killing vector field that, 
as well as having a well-defined continuum limit, is actually easier 
to construct than those of the Killing transport method, at least in our 
particular code.  Our method is extremely similar to that described 
by Cook and Whiting~\cite{Cook2007}, but was actually developed independently 
by one of the current authors~\cite{OwenThesis}.  

\subsection{Zero expansion, minimal shear}
Killing's equation,
\beq
\TwoGrad_{(A} \AKV_{B)} = 0,
\eeq
has two independent parts: the condition that $\vec \phi$ be expansion-free,
\beq
\AKVExpansion := \TwoMetric^{AB} \TwoGrad_A \AKV_B = 0,
\eeq
and the condition that it be shear-free,
\beq
\AKVShear_{AB} := \TwoGrad_{(A} \AKV_{B)} - \frac{1}{2} \TwoMetric_{AB} \AKVExpansion = 0,
\eeq
where uppercase latin letters index the tangent bundle to the 
two-dimensional surface, $\TwoMetric_{AB}$ is the metric on that surface, 
and $\TwoGrad_A$ is the torsion-free 
covariant derivative compatible with that metric.

When constructing 
approximate Killing vectors, a question arises: which condition is more 
important, zero expansion or zero shear?  Shear-free vector fields (conformal 
Killing vectors) are simply coordinate rotation generators in the common 
case of coordinate spheres in a conformally flat space.  They are therefore 
readily available in that context.  A very interesting and systematic approach 
to their use has been given by Korzynski~\cite{Korzynski2007}, and they have 
been used in the construction of conformally flat binary black hole initial 
data sets~\cite{Cook2004,Caudill-etal:2006}.  However, in the case of a 
general surface in a general spatial slice, 
the conformal Killing vectors are not known {\em a priori}, and they are more 
difficult 
to construct than expansion-free vector fields.  Expansion-free vector fields 
have the additional benefit of providing a gauge-invariant spin measure on 
a dynamical horizon~\cite{Ashtekar2003}\footnote{The dynamical horizon is 
essentially the world tube foliated by the apparent horizons.  The gauge 
freedom is that of extending the foliation off of this world tube.  The fact 
that this gauge invariance occurs when $\vec \phi$ is expansion-free can most 
easily be shown by expressing the factor $s^j K_{ij}$ in Eq.~\eqref{eq:Spin} in 
terms of the ingoing and outgoing null normals to the two-surface.  The boost 
freedom in these null normals has no effect on the spin when 
$\vec \phi$ is expansion-free.}, so we restrict attention to the 
expansion-free case.

Any smooth, expansion-free vector field tangent to a topological two-sphere 
can be written as
\beq
\phi^A = \epsilon^{AB} \TwoGrad_B z,\label{eq:expfree}
\eeq
where $\epsilon^{AB}$ is the Levi-Civita tensor and $z$ is some smooth 
potential function. 

We 
assume that the function $z$ has one local maximum, 
one local minimum, and no other critical points.  This is equivalent to the 
assumption that the orbits of $\vec \phi$ are simple closed loops.  In order 
for $\phi^A \phi_A$ to have the proper dimensions, $z$ must have dimensions of 
area.  For the case of the standard rotation generators of the metric 
two-sphere, the three $z$ functions are the three $\ell = 1$ spherical 
harmonics, multiplied by the square of the areal radius of the sphere.  

Within this space of expansion-free vector fields, 
we would now like to minimize the following positive-definite norm of the 
shear:
\beq
\ShearNorm := \oint_\AH \sigma_{BC} \sigma^{BC} \TwoAreaElement.
\eeq
Substituting Eq.~\eqref{eq:expfree} for $\vec \phi$  in this expression and 
integrating twice by parts, $\ShearNorm$ takes the form of an expectation 
value:
\beq
\ShearNorm = \oint_\AH z H z \TwoAreaElement , \label{eq:expectation_value}
\eeq
where $H$ is the self-adjoint fourth-order differential operator defined by
\beq
H z  = \TwoBiharmonic z + \OnAH{R} \TwoLaplacian z + \TwoGrad^A \OnAH{R} \medspace \TwoGrad_A z,
\eeq
and $\TwoLaplacian$ is the Laplacian on the (not necessarily round) sphere, 
$\TwoBiharmonic$ is its square, and $\OnAH{R}$ is 
the Ricci scalar curvature of the sphere.  
In our sign convention, $\OnAH{R} = 2$ on the unit sphere, so we 
can immediately see 
that $H z = 0$ when $z$ is an $\ell = 1$ spherical harmonic, and 
therefore that their associated vector fields are shear-free. 

It is now tempting to minimize the functional $\ShearNorm$ in 
\eqref{eq:expectation_value} with 
respect to $z$.  However, doing so will simply return the condition that $z$ 
lie in the kernel of $H$.  If there are no true Killing vectors, this will 
mean that $z$ is a constant, 
and therefore that $\vec \phi$ vanishes.  
We need to restrict the minimization 
procedure to cases that satisfy some normalization condition.  In this case, 
we require that the norm of the vector field,
\beq
\oint_\AH \phi^A \phi_A \TwoAreaElement \label{e:OurNorm},
\eeq
take some given positive value.  This restriction can be made with the use 
of a Lagrange multiplier.  Specifically, the functional we wish to minimize 
is
\begin{equation}
I[z] := \oint_\AH z H z \TwoAreaElement + \lambda \left(\oint_\AH \TwoGrad^A z \medspace \TwoGrad_A z \TwoAreaElement - N \right) \label{e:action}
\end{equation}
for some yet undetermined positive parameter $N$.  Note that $\lambda$ is the 
Lagrange multiplier and we have made use of the fact that 
Eq.~\eqref{eq:expfree} implies that 
$\vec \phi \cdot \vec \phi = \vec \TwoGrad z \cdot \vec \TwoGrad z$.  
Minimizing the functional $I$ with respect to $z$ returns a generalized 
eigenvalue problem:
\beq
H z = \lambda \TwoLaplacian z. \label{e:OurGenEigProb}
\eeq

It is at this point that we can most easily clarify the difference between 
our construction of approximate Killing vectors and that of Cook and Whiting 
in~\cite{Cook2007}.  The difference lies in the choice of norm 
in which the minimization problem is restricted.  Rather than fixing the 
norm \eqref{e:OurNorm} to take some fixed value in the minimization, Cook 
and Whiting instead fix the dimensionless norm:
\beq
\oint_\AH \OnAH{R} \phi^A \phi_A \TwoAreaElement. \label{e:CWNorm}
\eeq
In general, we see no particular reason to prefer either norm over the other, 
but for the current purposes we have at least an aesthetic preference for 
\eqref{e:OurNorm}, which is positive-definite even at high spin, whereas 
\eqref{e:CWNorm} is not, because the scalar curvature $\OnAH{R}$ of the 
horizon becomes negative near the poles at high spin.  If the 
norm \eqref{e:OurNorm} in Eq.~\eqref{e:action} is replaced by 
\eqref{e:CWNorm}, the result is the problem described in~\cite{Cook2007}:
\beq
H z = \lambda (\OnAH{R} \TwoLaplacian z + \TwoGrad^A \OnAH{R} \medspace \TwoGrad_A z). \label{e:CWGenEigProb}
\eeq

In our numerical code, we discretize~\eqref{e:OurGenEigProb} 
(or, optionally,~\eqref{e:CWGenEigProb}, but not for any results published 
here) and solve the resulting linear algebra problem 
with a LAPACK routine~\cite{lapack}.  Note, 
however, one technical peculiarity: the operators $H$ and $\TwoLaplacian$ 
in~\eqref{e:OurGenEigProb} share a 
kernel, the space of constant functions.  This means that this generalized 
eigenvalue problem is {\em singular}, a fact 
that can cause considerable difficulties for the numerical 
solution~\cite{LapackUsersGuide}.  The same can be said 
of~\eqref{e:CWGenEigProb}.  For our purposes, 
this complication is easily evaded.  Since we are working with a spectral 
code, it is easiest to discretize the problem using 
expansion into the spectral basis functions (coordinate spherical harmonics).  
When this is done, the space of constant functions---the shared kernel of 
the two operators---is simply the span of a single basis function: the 
constant, $Y_{00}$.  This basis function can easily be left out of the 
spectral expansion, and thereby removed from the numerical problem.  

Expansion into 
coordinate spherical harmonics has another practical advantage.  As noted 
earlier, for metric spheres in standard coordinates, the Killing 
vectors arise when $z$ is given by an $\ell = 1$ spherical harmonic.  
Thus, assuming our horizon is nearly round, and noticeably so in the given 
coordinates, the lowest basis functions (the $\ell = 1$ spherical harmonics) 
should nicely approximate the intended eigenfunctions.  The higher basis 
functions should simply provide small corrections.

In summary, the approach that we take to finding approximate Killing vectors 
begins with a spectral decomposition of Eq.~\eqref{e:OurGenEigProb}.  
This problem, of course, provides as many eigenvectors as there are elements 
of the spectral decomposition.  We restrict attention to the three 
eigenvectors with smallest eigenvalues (ignoring the vector corresponding to 
the constant eigenfunction, which is physically irrelevant and removed from 
discretization), as these are the ones corresponding 
to vector fields with the smallest shear, and at least for spheres that are 
only slightly deformed, the orbits of these vector fields are smooth closed 
loops.  

It must be noted that only the 
eigenvector with the smallest eigenvalue corresponds to a vector field with 
strictly minimum shear: even locally, all other eigenvectors are saddle points 
of the minimization problem.  The three of them taken together, however, 
provide a geometrically-defined subspace of the vector space of expansion-free 
vector fields, a natural generalization of the rotation generators on metric 
spheres.  Using these three vector fields (normalized as described in the 
next subsection), one can define ``components'' of 
the spin angular momentum of a black hole\footnote{In fact, using the higher 
eigenvectors, one could in principle compute higher-order multipole moments.  
We see this as a natural extension of the method laid out 
in~\cite{Schnetter2006} for defining the higher multipole moments of 
axisymmetric black holes.}, and from 
these components infer the spin around an arbitrary axis or even a spin 
``magnitude'' using a metric on this three-dimensional space of generalized 
rotation generators.  In practice, we have found no need to go quite so far.  
As mentioned in~\cite{Cook2007}, the approximate Killing vectors generally 
adapt themselves so well to the horizon that one of the components is much 
larger than the other two, so this is considered the spin 
magnitude, and the associated approximate Killing vector is considered to 
define the spin axis.  

\subsection{Normalization}
Solutions to the eigenproblem~\eqref{e:OurGenEigProb} can only determine 
the approximate Killing vectors up to a constant scaling.  Fixing this scaling 
is equivalent to fixing the value of $N$ in~\eqref{e:action}.  The standard 
rotation generators of metric spheres are normalized such that, when considered 
as differential operators along their various orbits, they differentiate with 
respect to a parameter that changes by a value of $2 \pi$ around each orbit.  
Naively one would like to fix the normalization of approximate Killing vectors 
in the same way, but a subtlety arises: we can only rescale the vector field 
by a fixed, constant value.  Rescaling differently along different orbits would 
introduce extraneous shear and would remove the vector field from the pure 
eigenspace of~\eqref{e:OurGenEigProb} in which it initially resided.  If an 
approximate Killing vector field has different parameter circumferences 
around different orbits, then it is impossible to rescale it such that the 
parameter distance is $2 \pi$ around every orbit.  The best one can ask is that 
$2 \pi$ is the {\em average} of the distances around the various orbits.  

To consider this in detail, introduce a coordinate system, topologically the 
same as the standard spherical coordinates on the metric sphere, but adapted 
to the potential function $z$ so that the latitude lines are the level surfaces 
of $z$ (and, in particular, the poles are at the two critical points we have 
assumed $z$ to have).  More precisely, choose $z$ for the zenith coordinate 
on the sphere, and an arbitrary rotational coordinate---say, the azimuthal 
angle in the encompassing spatial slice, describing rotations about the axis 
connecting the critical points of $z$---for the azimuthal coordinate $\varphi$ 
on the sphere.  If the parameter $\tau$ is defined such 
that $\vec \phi = (d/d \tau)_{z = {\rm const.}}$, then in the basis related to 
these coordinates, the components of $\vec \phi$ are:
\begin{eqnarray}
\phi^z(z, \varphi) &=& \left(\frac{dz}{d\tau}\right)_{z={\rm const.}} = 0,\\
\phi^\varphi(z, \varphi) &=& \left(\frac{d\varphi}{d\tau}\right)_{z={\rm const.}}.
\end{eqnarray}
Around a closed orbit ${\cal C}(z)$, at fixed $z$, the parameter $\tau$ 
changes by a value of:
\beqa
\tau(z) &=& \int_{{\cal C}(z)} \frac{d\varphi}{\phi^\varphi(z,\varphi)}
\label{eq:tau1}\\
&=& \int_{{\cal C}(z)} \frac{d\varphi}{\epsilon^{\varphi z} \partial_z z}
\label{eq:tau2}\\
&=& \int_{{\cal C}(z)} \sqrt{\TwoMetric} d\varphi,\label{eq:tau3}
\eeqa
where $\TwoMetric$ is the determinant of the surface metric, evaluated in the 
$(z, \varphi)$ coordinates.  Note that Eq.~\eqref{eq:tau3} follows from 
Eq.~\eqref{eq:tau2} by the fact that the condition 
$\TwoMetric_{AB}\TwoMetric_{CD}\epsilon^{AC}\epsilon^{BD} = 2$ implies 
$\epsilon^{\varphi z} = 1/\sqrt{\TwoMetric}$.  The average value of $\tau$, over the various 
orbits, is:
\beqa
\left< \tau \right> &=& \frac{1}{z_{\rm max} - z_{\rm min}} \int_{z_{\rm min}}^{z_{\rm max}} \int_{{\cal C}(z)} \sqrt{\TwoMetric} d\varphi dz \\
&=& \frac{A}{z_{\rm max} - z_{\rm min}},
\eeqa
where $A$ is the surface area of the apparent horizon.  Requiring this average 
to equal $2 \pi$, we arrive at the normalization condition:
\begin{equation}
2 \pi (z_{\rm max} - z_{\rm min}) = A. \label{e:normalization}
\end{equation}

This normalization condition requires finding the minimum and maximum values 
of the function $z$, 
which is only computed on a discrete grid.  In our spectral code, in 
particular, this numerical grid is quite coarse, so numerical interpolation is 
needed, in combination with an optimization routine.  We have implemented such 
routines to search for $z_{\rm min}$ and $z_{\rm max}$, but a 
numerically-cheaper 
normalization condition would be of interest.  Such a condition arises 
when one assumes that the black hole under consideration is approximately 
Kerr.  In the Kerr metric, for the function $z$ generating the true rotation 
generator of the Kerr horizon, the following identity holds:
\beq
\oint_{\AH} \left(z - \left<\left< z \right>\right> \right)^2 \TwoAreaElement = \frac{\Area^3}{48 \pi^2}, \label{e:alt_normalization}
\eeq
where $\left<\left< z \right>\right>$ is the average of $z$ over the sphere.  
The existence of an identity of this form is somewhat nontrivial: the 
fact that the right side is given purely by the horizon area, and that it does 
not involve the spin of the Kerr hole, is what makes this identity useful as a 
normalization condition.  This normalization is much easier to impose, and 
requires significantly less numerical effort.  

\begin{figure}
\includegraphics[scale=0.3]{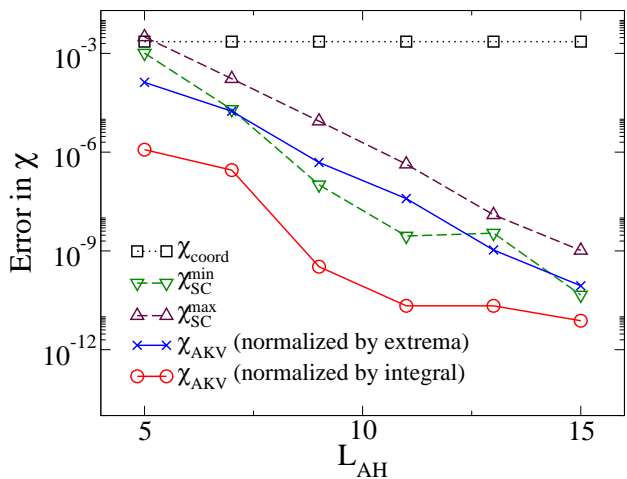}
\caption{ \label{fig:DefKerrConv} Error, relative to the analytic solution, of 
the spin on the horizon of a Kerr black hole in slightly deformed coordinates. 
The vertical axis represents $|\SMM_{\rm computed} - \SMM_{\rm analytic}|$, 
and data are shown for the spin computed with the standard coordinate rotation 
vector (in deformed coordinates, so not a true Killing vector), and with our 
approximate Killing vectors 
(AKV) using both the extremum norm, Eq.~\eqref{e:normalization}, and the 
integral norm, Eq.~\eqref{e:alt_normalization}.  The spin computed from the 
coordinate rotation vector quickly converges to a physically inaccurate 
result.  The spin from approximate Killing vectors converges in resolution 
$L_{\rm AH}$ to the correct value $\SMM = 1/2$.  Curves are also shown for the 
two spin measures defined in the Appendix~\ref{sec:SpinFromShape}.  These spin 
measures also converge exponentially to the physically correct result.
}
\end{figure}

To close the discussion of spin computed from approximate Killing vectors, we 
demonstrate the effectiveness of the method in a simple test case: 
an analytic Kerr black hole in slightly deformed coordinates.  We begin with a 
Kerr black hole of dimensionless spin parameter $\SMM = 1/2$, in Kerr-Schild 
coordinates, but we rescale the $x$-axis by a factor of $1.1$.  This 
rescaling of the $x$ coordinate causes the coordinate rotation vector 
$x \partial_y - y \partial_x$ to no longer be the true, geometrical rotation 
generator.  And indeed, when we compute the quasilocal angular 
momentum~\eqref{eq:Spin} on the horizon using this coordinate vector, the 
result converges to a physically inaccurate value, as demonstrated by the 
black dotted curve in Fig.~\ref{fig:DefKerrConv}.  If, however, the 
approximate Killing vectors described above are used, the result is not only 
convergent, but physically accurate.  Because the accuracy is slightly better 
with the normalization condition of Eq.~\eqref{e:alt_normalization}, that is 
the condition we use for all results presented in this paper.

%%%%%%%%%%%%%%%%%%%%%%%%%%%%%%%%%%%%%%%%%%%%%%%%%%%%%%%%%%%%%%%%
%%%%%%%%%%%%%%%%%%%%%%%%%%%%%%%%%%%%%%%%%%%%%%%%%%%%%%%%%%%%%%%%

\section{Scalar-curvature spin (SC spin)}
\label{sec:SpinFromShape}

In this appendix, we define a spin measure in terms of the intrinsic
geometry of the horizon, which we compare with the AKV spin in
Sec.~\ref{sec:Evolutions}.
The AKV spin described in
Appendix~\ref{sec:QuasilocalSpin} 
is a well-defined measure 
of black hole spin, even when the holes' horizons
have only approximate symmetries. At times sufficiently 
before or after the holes merge, 
however, the horizons will not be too tidally distorted 
and thus
will not be too different from the exactly-axisymmetric horizons of 
Kerr black holes.

By assuming that the geometric properties of the 
horizon behave 
precisely as they do for a Kerr black hole, 
one can infer the hole's spin from those properties. 
For instance, it is common to measure polar and equatorial 
circumferences of the apparent horizon; the spin is then obtained 
by finding the Kerr spacetime with the same 
circumferences~\cite{AnninosEtAl:1994,BrandtSeidel:1995,AlcubierreEtAl:2005}. 

To avoid introducing coordinate dependence by defining ``polar'' and 
``equatorial'' planes, we infer the spin from the 
horizon's intrinsic scalar curvature~$\HorizonShape$.
The horizon scalar curvature $\HorizonShape$ 
has previously been studied analytically for 
Kerr-Newman black holes~\cite{Smarr:1973} and for
Kerr black holes perturbed by a distant moon~\cite{Hartle:1974}. 
Numerical studies of $\HorizonShape$ have focused attention on the 
quasinormal ringing of single, perturbed,
black holes~\cite{AnninosEtAl:1994} as well as on the shape 
of the individual and common event horizons in 
Misner~data~\cite{MassoEtAl:1999}. 
To our knowledge, the 
scalar curvature $\HorizonShape$ 
has not been previously used to infer the horizon 
spin in numerical simulations.

At a given point on a Kerr black hole's horizon, 
the horizon scalar curvature  $\HorizonShape$ depends only on 
the hole's mass $\HorizonMass$ and 
spin $\Spin$. 
The extrema of $\HorizonShape$ can be expressed 
in terms of
  the irreducible mass and dimensionless spin of the Kerr black hole 
via Eqs.~(\ref{eq:SMMDef})--(\ref{eq:GoodSpinNorm}) as 
\begin{subequations}
\begin{eqnarray}
\ShapeMin & = & \frac{-1 + 2 \sqrt{1-\SMM^2}}{2 \Mirr^2},
\label{eq:SpinFromShapeMin}\\
\ShapeMax & = & -\frac{2}{\Mirr^2 \SMM^4} 
\left(-2 + \SMM^2 + 2 \sqrt{1-\SMM^2}\right).\label{eq:SpinFromShapeMax}
\end{eqnarray}
\end{subequations} 

Solving for $\SMM$ and requiring it to be real 
yields $\SMM$ as a function of 
$\Mirr$ and either $\ShapeMin$ or $\ShapeMax$. 
We take these functions as {\em definitions} of the spin,
even when the space-time is not precisely Kerr:
\begin{subequations}
\begin{eqnarray}\label{eq:SpinMin}
%% |\SpinFromShapeMax| := & & \nonumber\\ 
%% \frac{1}{2}\sqrt{3-4\Mirr^2 \ShapeMin 
%% \left(1+\Mirr^2 \ShapeMin\right)}, & & \\
\left(\SpinFromShapeMin\right)^2 & := & 
1 - \left[\frac{1}{2}+\Mirr^2 \ShapeMin\right]^2, \\
%% \SpinFromShapeMax^2 := & & \nonumber\\ 
%% \frac{\sqrt{2}}{\Mirr \sqrt{\ShapeMax}} 
%% \sqrt{-1+\sqrt{2} \Mirr \sqrt{\ShapeMax}}. & &
\left(\SpinFromShapeMax\right)^2 & := & 
\frac{-2+2\sqrt{2 \Mirr^2 \ShapeMax}}{\Mirr^2 \ShapeMax}
\label{eq:SpinMax}
\end{eqnarray}
\end{subequations} The definitions of the spin given by 
Eqs.~(\ref{eq:SpinMin})--(\ref{eq:SpinMax}) are 
manifestly independent of spatial
coordinates
and are well-defined
for black holes that are tidally deformed. Also, as they only involve the 
intrinsic two-dimensional geometry of the apparent horizon, they are also 
manifestly independent of boost gauge, in the sense described in 
the previous appendix.  

We expect $\SpinFromShapeMin$ and $\SpinFromShapeMax$ 
to be reasonable measures 
only if tidal forces can be neglected. 
Tidal forces scale with the cube of the separation of the holes; for binary 
with holes of equal mass $\HorizonMass$ 
and separation $\CoordSep$, tidal coupling is 
negligible when $\ShapeMax - \ShapeMin \gg \HorizonMass/\CoordSep^3$.

We find it convenient to 
compute $\OnAH{R}$ from i) the 
scalar curvature $\SRicciS$ associated with the three-dimensional 
metric $\SMetric_{ij}$ of the spatial slice $\Slice$, and ii) 
the outward-pointing unit-vector field 
$\SSpatialNormal^i$ that is 
normal to $\AH$. This can by done by means of Gauss's equation 
[e.g., Eq.~(D.51) of Ref.~\cite{carrollTextbook} (note that
the Riemann tensor in Ref.~\cite{carrollTextbook} 
disagrees with ours by an overall sign)]
\begin{eqnarray}\label{eq:GaussEqn}
%\OnAH{R} = R - 2 R_{ij} s^i s^j - \OnAH{K}^2 + \OnAH{K}^{ij} \OnAH{K}_{ij},
\HorizonShape = \SRicciS - 2 \SRicci_{ij} \SSpatialNormal^i \SSpatialNormal^j - \OnAH{\ExCurv}^2 + \OnAH{\ExCurv}^{ij} \OnAH{\ExCurv}_{ij},
\end{eqnarray} 
where $\SRicci_{ij}$ and $\SRicciS$ were defined after Eq.~(\ref{eq:Mom}), and
where $\OnAH{\ExCurv}_{ij}$ denotes the extrinsic curvature 
of the the apparent horizon $\AH$ embedded in $\Slice$ 
(not to be confused with $\ExCurv_{ij}$,
the extrinsic curvature of the slice $\Slice$ embedded in ${\cal M}$).
The horizon extrinsic curvature is given by
\begin{eqnarray}\label{eq:AHExCurvDef}
%\OnAH{K}_{ij} = \nabla_i s_j - s_i s^k \nabla_k s_j.
\OnAH{\ExCurv}_{ij} = \SCD_i \SSpatialNormal_j - \SSpatialNormal_i \SSpatialNormal^k \SCD_k \SSpatialNormal_j.
\end{eqnarray} Inserting Eq.~(\ref{eq:AHExCurvDef}) 
into Eq.~(\ref{eq:GaussEqn}) shows that $\HorizonShape$ can be evaluated 
exclusively in terms of quantities defined on the 
three-dimensional spatial slice $\Slice$.

The accuracy of these spin measures is demonstrated in 
Fig.~\ref{fig:DefKerrConv}, which shows a Kerr black hole 
with $\SMM = 1/2$ in slightly 
deformed coordinates so that the coordinate rotation vector no longer 
generates a symmetry.  Again, both $\SpinFromShapeMin$ and $\SpinFromShapeMax$ 
converge exponentially to the physically accurate result.

%%%%%%%%%%%%%%%%%%%%%%%%%%%%%%%%%%%%%%%%%%%%%%%%%%

%%%%%%%%%%%%%%%%%%%%%%%%%%%%%%%%%%%%%%%%%%%%%%%%%%%%%%%%%%%%%%%%%%%%%%%%%%%%%%%
%\section*{References}
%%%%%%%%%%%%%%%%%%%%%%%%%%%%%%%%%%%%%%%%%%%%%%%%%%%%%%%%%%%%%%%%%%%%%%%%%%%%%%%
\bibliography{References/References}

\end{document}